\documentclass[]{pasj01}
\usepackage{subfigure}
\usepackage{lscape}

\Received{2020/07/30}
\Accepted{2021/04/19}

\begin{document}

\title{
Unification of BL Lac objects, FR I and FR II(G) radio galaxies and Doppler factor Estimation for BL Lac Objects}

\author{Xu-Hong \textsc{Ye}\altaffilmark{1,2,3}}
\author{Jun-Hui \textsc{Fan}\altaffilmark{1,2,3}}
\altaffiltext{1}{Center for Astrophysics, Guangzhou University, Guangzhou 510006, China}
\altaffiltext{2}{Astronomy Science and Technology Research Laboratory of Department of Education of Guangdong Province, Guangzhou 510006, China}
\altaffiltext{3}{Key Laboratory for Astronomical Observation and Technology of Guangzhou, Guangzhou 510006, China}
\email{fjh@gzhu.edu.cn}


\maketitle

\begin{abstract}
In this work, we collected a sample of BL Lacs, FR I and FR II(G) radio galaxies with available core and extended emissions from published works to discuss the unified schemes and estimate the Doppler factor for BL Lacs. Wilcoxon rank-sum test and Kolmogorov-Smirnov test both suggest that the probabilities for the distribution of  the extended luminosity of BL Lacs and that of  FR I and FR II(G) radio galaxies to be from the same parent distribution are $p_{\rm{WRS}}=0.779$ and  $p_{\rm{K-S}}=0.326$, suggesting they are unified. Based on this unified schemes, we propose to estimate the Doppler factors for BL Lacs. Comparing the Doppler factor estimated by the fitting/regression method with those for the common sources in the literatures, we found a good linear correlation for common sources.
\end{abstract}

\section{Introduction}
Active Galactic Nuclei (AGNs) are a special class of galaxies, showing extreme observation properties. Blazars are the extreme class of AGNs, which can be subdivided into two subclasses because of their behavior of emission lines: BL Lacertae objects (BL Lacs) with weak or no emission line and flat-spectrum radio quasars (FSRQs) with strong ones. BL Lacs show some extreme observation properties such as rapid and high amplitude variation, high and variable polarization, very weak or no emission line features, superluminal motions, or even high energy $\gamma$-ray emissions \citep{sti91,fan99,fan02,xiao19}. When the viewing angle between relativistic jet and line of sight is small, a Doppler beaming effect should be taken into considered. The observed flux density $S^{\rm{ob}}$ is enhanced by the Doppler factor, $S^{\rm{ob}}=S^{\rm{in}}\delta^q$, where $S^{\rm{in}}$ is the intrinsic flux density, and $\delta$ is a Doppler factor, $q=2+\alpha$ for a continuous jet, $q=3+\alpha$ for a spherical jet \citep{sch79}, in which $\alpha$ is the spectral index ($S_{\nu}\propto v^{-\alpha}$, $S_{\nu}$ for the monochromatic flux density). The Doppler factor is important for emissions in BL Lacs, which is determined by velocity ($\beta$) and the viewing angle ($\theta$): $\delta=[\Gamma(1-\beta \cos \theta)]^{-1}$, where $\Gamma$ is the Lorentz factor, satisfying $\Gamma=1 / \sqrt{1-\beta^{2}}$. 

Some methods were proposed to estimate the Doppler factor \citep{ghi93,mat93,lah99,fan13,chen18,lioda18,zhang20}. \citet{ghi93} adopted  the synchrotron self-Compton (SSC) model and assumed the synchrotron high frequency cutoff as $10^{14}$Hz to obtain the Doppler factors. Their results show that the Doppler factors are the largest for core-dominated quasars, intermediate for BL Lacs, and the smallest for radio galaxies and lobe-dominated quasars. \citet{lah99} adopted a method using the total radio flux density variations to assess Doppler factors and found high-polarization quasars with higher Doppler factors, while BL Lacs and low-polarization quasars have smaller Doppler factors. Following \citet{mat93}, \citet{fan13} assumed the $\gamma$-ray timescale as one day and calculated the lower limit for $\gamma$-ray Doppler factors from X-ray and $\gamma$-ray emissions.  Based on the spectral energy distributions (SEDs) model, \citet{chen18} estimated the Doppler factors for a sample of 999 blazars. The Doppler factor of BL Lacs is significantly larger than that of FSRQs in the work by \citet{chen18}. \citet{lioda18} analyzed and constrained the average equipartition brightness temperature ($T_{eq}$) of the whole sample same as those ($T_{eq}=2.78\times10^{11}K$) of FSRQs to obtain Doppler factors for a larger sample of blazars, showing that blazars are strongly beamed sources with higher Doppler factor on average than that of radio galaxies and unclassified sources. \citet{zhang20} proposed a new method based on the correlation between $\gamma$-ray luminosity and broad-line region luminosity to estimate the Doppler factors for a sample of 350 blazars. We can see that different method gives different Doppler factor value because of the different assumption in the literatures. For instance,  based on the SSC model, \citet{ghi93} obtained a Doppler factor of $\delta = 2.1$ for the BL Lac 0716+714, while \citet{hova09} obtained  $\delta = $ 10.9 using the radio variation and \citet{lioda18} derived a Doppler factor $\delta$ = 31.3 by adopting the same intrinsic brightness temperature for all the samples.

Radio galaxies are also a subclass of AGNs. From a work by \citet{fana74}, the radio galaxies are classified into two types, based on their luminosity and the morphology. For some sources, the radio luminosity is mainly from the central part, this radio galaxy is classified as class I, called as Fanaroff-Riley class I (FR I) radio galaxy; some sources that their radio luminosity is mainly from the outer edge of the galaxy, this type of radio galaxy is regarded as class II radio galaxies, and called as Fanaroff-Riley class II (FR II) radio galaxy \citep{fana74}. The centeral compact core emission in FR I radio galaxies (hereafter FR Is) is found to be identified with strongly associated with optical synchrotron radiation, which is produced in the inner regions of the relativistic jet \citep{chi99}. In addition, \citet{ver02} had analyzed a complete sample of 21 nearby FR Is and proposed that the radio and optical core emissions of these samples are likely synchrotron radiation from inner jet because (a) radio and optical core emission are closely correlated, (b) the radio to optical spectral indices are similar to those for extended optical jets, (c) there is a suggestive trend with independent estimates from jet orientation, (d) the residual for radio-H$\alpha+$[N \uppercase\expandafter{\romannumeral2}] core correlation and that for optical-H$\alpha+$[N \uppercase\expandafter{\romannumeral2}] core correlation are well correlated with each other. From the optical spectrum, FR II radio galaxy can be subdivided into  two subclass: FR II(G) and FR II(Q). The FR II(G) radio galaxy shows  its optical type as that of a galaxy, while the spectral type of FR II(Q) resemble an optical type of a quasar \citep{xie93}. 

In the unification of AGNs, the viewing angle is invoked to explain the different observation properties of AGNs. It  was proposed that BL Lacs and FR Is were unified, in this unified scheme, the FR Is are the parent population of the BL Lacs \citep{urry91, ghi93, urry95}. \citet{urry91} discussed the unification of BL Lacs and FR Is by computing luminosity functions, and found this samples were consistent with the beaming hypothesis that the BL Lacs are FR Is seen face on. \citet{ghi93} computed the average Lorentz factor ($\Gamma=8.32$)  and viewing angle ($\theta=11.8$) for a sample of BL Lacs, which are consistent with other results from the attempts to unify FR I/BL Lac schemes, suggesting that the FR Is should be the parent population of the BL Lacs.  Meanwhile, some authors suggested that FR II radio galaxies may be a part of parent population for BL Lacs \citep{xie93,owen96,fan97}. \citet{xie93} discussed the unified schemes of 75 BL Lacs, 27 FR I and 45 FR II(G) radio galaxies (hereafter FR II(G)s) using the Hubble diagram, and found BL Lacs and FR Is and FR II(G)s fit the same Hubble relation very well, supporting the unified schemes of BL Lacs and FR Is should include the FR II(G)s.

Due to the unobservable characteristics, it is difficult to obtain the Doppler factor. The previous studies mentioned above  used different hypothesis to assess the Doppler factor \citep{ghi93, mat93, lah99,fan13,chen18,lioda18,zhang20}.  In this sense, any method estimating the Doppler factor is important for AGN studying. In a two-component model \citep{urry84},  the core with a relativistic jet plus extended component is considered, in which the core emission is enhanced by the Doppler beaming effect, while the extended component has no beaming effect and shows their intrinsic emission. Since the extended emissions are unbeamed, then we can discuss the relationship of the extended luminosity for BL Lacs and FR I with FR II(G) radio galaxies (hereafter FR I/II(G)s). If BL Lacs and FR I/II(G)s are unified with FR I/II(G)s being the parent population of BL Lacs, then based on this unified schemes, we assumed that the intrinsic core and extended luminosity of BL Lacs should follow the same correlation as that of FR I/II(G)s, then we can estimate the Doppler factor of BL Lacs by using the core to extended luminosity fitting/regression method or ratio method of FR I/II(G)s. The structure of this work is arranged as follows. In Sect. 2, the unified schemes for BL Lacs and FR I/II(G)s is discussed using Wilcoxon rank-sum (WRS) test and Kolmogorov-Smirnov (K-S) test. In Sect. 3, based on this unified model, we estimate the Doppler factor. In Sect. 4, some discussions are presented, and some conclusions are shown in Sect. 5.

\section{Samples and unified schemes}
\subsection{Samples}
In this work, we collected 297 BL Lacs, 87 FR Is and 41 FR II(G)s with core and extended fluxes or luminosities at 5 GHz from the literature and showed them in Table \ref{tab1}, 
\setlength{\tabcolsep}{3pt}
\begin{table*}
\caption{The core and extended fluxes or luminosities of whole samples and the Doppler factor of BL Lacs.}
\begin{tabular}{@{}lcccccccccccccc@{}}
\hline
Source&class&$z$ & $S_{\rm{core}}$ & $S_{\rm{ext}}$&Ref&$logL_{\rm{core}}$&$logL_{\rm{ext}}$&Ref&$\delta_2$ &$\delta_2		$&$\delta_2$&$\delta_3$&$\delta_3$&$\delta_3$ \\
		 Name& & &(mJy)&(mJy)& &(W\ Hz$^{-1})$&(W\ Hz$^{-1})$  & &($\alpha_c=0$ &0.5&-0.5)&($\alpha_c=0$&0.5&-0.5) \\
		(1)&(2)&\,(3)&(4)&\,\,(5)&(6)&(7)&(8)&(9)&(10)&(11)&(12)&(13)&(14)&(15)\\
\hline
		0003+003	&	B	&	1.037	&	480	&	206	&	P19	&	27.17	&	26.81	&		&	18.74	&	10.43	&	49.79 	&	7.06	&	5.34	&	10.43 	\\
		0003-066	&	B	&	0.347	&	1850	&	1504.47	&	P19	&	26.73	&	26.64	&		&	12.55	&	7.57	&	29.16 	&	5.40	&	4.24	&	7.57 	\\
		0007+124	&	I	&	0.156	&	4	&	751.02	&	F11	&	23.47	&	25.75	&		&		&		&		&		&		&		\\
		0007+472	&	B	&	0.280	&	67	&	22	&	P19	&	25.17	&	24.69	&		&	7.69	&	5.11	&	15.17 	&	3.90	&	3.21	&	5.11 	\\
		0011+1853	&	B	&	0.473	&	140	&	104	&	P19	&	25.89	&	25.76	&		&	8.58	&	5.58	&	17.57 	&	4.19	&	3.42	&	5.58 	\\
		0013+790	&	II(G)	&	0.840	&	4.4	&	1028.6	&	F11	&	25.21	&	27.58	&		&		&		&		&		&		&		\\
		0021+055	&	B	&	2.050	&	28	&	53	&	P19	&	26.56	&	26.83	&		&	9.04	&	5.82	&	18.84 	&	4.34	&	3.52	&	5.82 	\\
		0029-271	&	B	&	0.333	&	11	&	105	&	P19	&	24.67	&	25.65	&		&	2.27	&	1.93	&	2.98 	&	1.73	&	1.60	&	1.93 	\\
		0032+595	&	B	&	0.086	&	44	&	5	&	P19	&	23.92	&	22.97	&		&	5.71	&	4.03	&	10.20 	&	3.19	&	2.71	&	4.03 	\\
		0033+156	&	B	&	1.162	&	125	&	28	&	P19	&	26.96	&	26.31	&		&	20.43	&	11.18	&	55.86 	&	7.47	&	5.61	&	11.18 	\\
		0038+328	&	II(G)	&	0.482	&	0.47	&	1199.53	&	NED	&	23.64	&	27.05	&		&		&		&		&		&		&		\\
		0039+398	&	I	&	0.109	&	1	&	225	&	P19	&	22.52	&	24.88	&		&		&		&		&		&		&		\\
		0043+008	&	B	&	2.149	&	2	&	2	&	P19	&	25.27	&	25.27	&		&	5.86	&	4.11	&	10.55 	&	3.25	&	2.75	&	4.11 	\\
		0044+193	&	B	&	0.181	&	7	&	17	&	P19	&	23.71	&	24.10	&		&	2.13	&	1.83	&	2.74 	&	1.65	&	1.54	&	1.83 	\\
		0048-09	&	B	&	0.634	&	887	&	108.45	&	F11	&	26.98	&	26.07	&		&	24.62	&	12.97	&	71.62 	&	8.46	&	6.24	&	12.97 	\\
		0052+251	&	B	&	0.154	&	1	&	1	&	P19	&	22.73	&	22.73	&		&	1.71	&	1.54	&	2.05 	&	1.43	&	1.36	&	1.54 	\\
		0053+260	&	I	&	0.195	&		&		&		&	23.00	&	25.04	&	F11	&		&		&		&		&		&		\\
		$\cdots$ & $\cdots$ & $\cdots$ & $\cdots$ & $\cdots$ & $\cdots$ & $\cdots$ & $\cdots$ & $\cdots$ & $\cdots$ & $\cdots$ & $\cdots$ & $\cdots$ & $\cdots$ & $\cdots$\\
\hline
\end{tabular}\label{tab1}
\medskip
\tabnote{Note: Col. 1 gives the source name, Col. 2 the classification, B for BL Lacs, I for FR I, II(G) for FR II(G),  Col. 3 redshift, Col. 4 the core flux density at 5 GHz, Col. 5 the extended flux density at 5 GHz, Col. 6 the references for Col. 4 and 5,  NED for NASA/IPAC Extragalactic Datebase, M93 for \citet{mor93}, PS93 for \citet{per93},  K05 for \citet{kov05}, K10 for \citet{kharb10},  F11 for \citet{fan11}, DM14 for \citet{mauro14}, P19 for \citet{pei19}, Col. 7 the core luminosity at 5 GHz, Col. 8 the extended luminosity at 5 GHz, Col.9 the references for Col. 7 and 8, Z95 for \citet{zir95},  B11 for \citet{bro11},  Col. 10 the Doppler factor for $q=2$, $\alpha_c=0$, Col. 11 the Doppler factor for q=2, $\alpha_c=0.5$,  Col. 12 the Doppler factor for $q=2$, $\alpha_c=-0.5$, Col. 13 the Doppler factor for $q=3$, $\alpha_c=0$, Col. 14 the Doppler factor for $q=3$, $\alpha_c=0.5$, Col. 15 the Doppler factor for $q=3$, $\alpha_c=-0.5$.

( A portion is shown here for guidance regarding its form and content. The Table 1 is  published in its entirety in Appendix.)}
\end{table*}
in which the core and extended fluxes at 5 GHz are listed in Col. 3-4 with their references in Col. 5, and their corresponding luminosities  are  listed in Col. 7-8. For some sources, their core and extended luminosities are available in the literatures \citep{zir95, bro11, fan11} as listed in Col. 7-9. Generally, the measured frequency of data is different in different literature.  Because most of the measured radio frequency is at 5 GHz, \citet{fan11} and \citet{pei19} transferred the data at other measured frequency to 5 GHz, $S^{5 GHz}_c=S^{\nu,obs}_c$, $S_{e x t}^{5 G H z}=S_{\rm{e x t}}^{\nu, o b s}(\frac{\nu}{5 G H z})^{\alpha_{\rm{e x t}}}$, with $\alpha_{\rm{ext}}=0.75$ and $\alpha_c=0.0$ \citep{fan11,pei19}.

We compare BL Lacs in the present sample (BL Lacs$_{\rm{TW}}$, 297 sources) with BL Lacs (BL Lacs$_{\rm{ref}}$, 649 sources) in the references \citep{fan11,pei19} and Roma-BZCAT \citep{mas15}, to discuss the completeness of BL Lacs of the sample. The redshift of BL Lacs$_{\rm{TW}}$ is in the range of 0.026 to 3.2, and the coverage range for redshift distribution of BL Lacs$_{\rm{TW}}$ is similar to that of BL Lacs$_{\rm{ref}}$, 
\begin{figure}[bht]
\includegraphics[width=1.1\columnwidth]{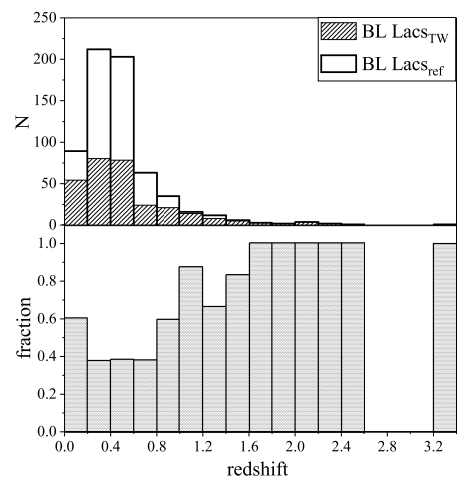}
\caption{The upper panel is the redshift distributions for BL Lacs, the unhatched area for BL Lacs (BL Lacs$_{\rm{ref}}$) in the references \citep{fan11,pei19} and Roma-BZCAT \citep{mas15}, the hatching area for BL Lacs in the present sample (BL Lacs$_{\rm{TW}})$. The lower panel is the fraction between BL Lacs$_{\rm{TW}}$ and BL Lacs$_{\rm{ref}}$ in different redshift bin.} 
\label{fig.BLredshift}
\end{figure}
which is also in the range of  0.026 to 3.2, as shown in the upper panel of Fig. \ref{fig.BLredshift}. Additionally, the fraction between BL Lacs$_{\rm{TW}}$ and BL Lacs$_{\rm{ref}}$ is obtained in different redshift bin and presented in the lower panel of Fig. \ref{fig.BLredshift}. About half of BL Lacs$_{\rm{ref}}$ within redshift $<$ 1 and almost all the BL Lacs$_{\rm{ref}}$ with redshift $>$ 1 were selected for research. 
FR I/IIs in the present sample (FRs$_{\rm{TW}}$, 128 sources) are also compared with FR I/IIs (FRs$_{\rm{ref}}$, 395 sources) in the references \citep{mat93, zir95, bro11, fan11, pei19}. The redshift distribution of FRs$_{\rm{TW}}$ is in a range of 0.003 to 1.132, and that of FRs$_{\rm{ref}}$ ranges from 0.002 to 2.009, as shown in upper panel of Fig. \ref{fig.FRredshift}. The fraction between FRs$_{\rm{TW}}$ and FRs$_{\rm{ref}}$ in different redshift bin is also presented in the lower panel of Fig. \ref{fig.FRredshift}.
One can see that the present sample has a smaller redshift than that from the references. We think that the reason is because the radio galaxy with higher redshift is weak so that it is hard for one to obtain the core and the extended emissions, therefore we can only separate the core and the extended emissions for the low redshift radio galaxy. We hope that we can obtain more core and extended emissions with higher redshift in the future.
\begin{figure}[bht]
\includegraphics[width=1.1\columnwidth]{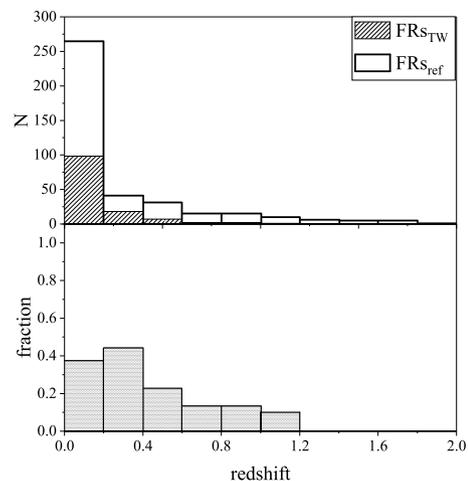}
\caption{The upper panel is the redshift distributions for FR I/IIs, the unhatched area for FR I/IIs (FRs$_{\rm{ref}}$) in the references \citep{mat93, zir95, bro11, fan11, pei19}, the hatching area for FR I/IIs in the present sample (FRs$_{\rm{TW}})$. The lower panel is the fraction between FRs$_{\rm{TW}}$ and FRs$_{\rm{ref}}$ in different redshift bin.}
\label{fig.FRredshift}
\end{figure}

\subsection{ Luminosity Calculation}

From the core and extended fluxes, one can calculate the corresponding luminosities, $L=4\pi d_L^2S_{\nu}$, where $d_L$ is a luminosity distance expressed by
 \begin{equation}
d_L=(1+z)\frac{ c}{H_{0}} \int_{1}^{1+z} \frac{1}{\sqrt{\Omega_{M} x^{3}+1-\Omega_ {M}}} dx 
\end{equation}
with $\Omega_{\Lambda}\sim0.692$, $\Omega_M\sim0.308$ and $H_0=67.8$km s$^{-1}$ Mpc$^{-1}$ \citep{pla16}.

For the 297 BL Lacs, our calculation shows that the logarithm of the core luminosity, log$L_{\rm{core}}$, is in a range of  
 $\log L_{\rm{core}}$ = 21.96 $--$ 28.70 with an average of  25.53, where $L_{\rm{core}}$ indicates the core luminosity
in the unit of W Hz$^{-1}$, and that of the extended luminosity, log$ L_{\rm{ext}}$, is in a range of $\log L_{\rm{ext}}$ = 21.56 $--$ 28.19, where $L_{\rm{ext}}$ indicates the extended luminosity
in the unit of W Hz$^{-1}$.

For the 87 FR Is, we have $\log L_{\rm{core}}$ = 20.90 $--$ 25.40 with an average of 23.32 for the core luminosity, and  $\log L_{\rm{ext}}$ = 22.49 $--$ 26.80 with an average of 24.43 for the extended luminosity. While for the 41 FR II(G)s,  the core luminosity is found to be in a range from $\log L_{\rm{core}}$ = 22.69 $--$ 26.56 with an average of 24.43  and the extended luminosity in a range of  $\log L_{\rm{ext}}$ = 25.14 $--$ 27.85 with an average of 26.35.

\subsection{Unification Scheme}

From the relativistic two-component model \citep{urry84}, the total luminosity consists of the core (beamed) luminosity and extended (unbeamed) luminosity, in which core luminosity is enhanced by the  relativistic beaming effect, while the extended luminosity displays its intrinsic value. In order to investigate the unification between BL Lacs and FR I/II(G)s, we used the extended luminosity for these samples. Because of the FR I/FR II dichotomy from the \citet{fana74},  the different property in the luminosity between the FR I and FR II is widely accepted for decades. FR Is show their behavior as low radio power, and FR IIs behave as high radio power owing to their different jet effect. In our samples, the FR II(G)s have a higher  extended luminosity 
\begin{figure}[bht]
\includegraphics[width=1\columnwidth]{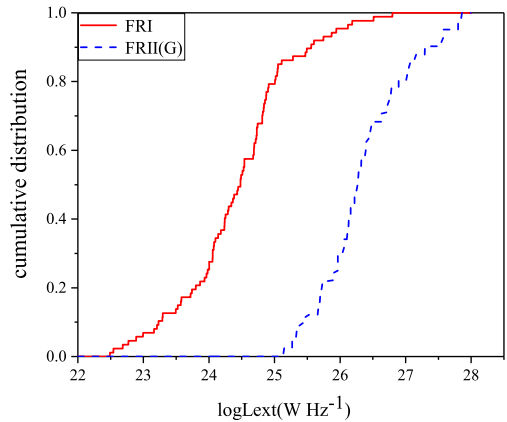}
\caption{The cumulative distributions for extended luminosity of FR Is and FR II(G)s, in which the solid line represents FR Is and broken line represents FR II(G)s.}
\label{fig.FRcum}
\end{figure} 
ranging from $\log L_{\rm{ext}}$ = 25.14 $--$ 27.85, but FR Is have lower extended luminosity,  $\log L_{\rm{ext}}$ = 22.46 $--$ 26.80. The cumulative distributions for extended luminosity of FR Is and FR II(G)s are presented in Fig. \ref{fig.FRcum}.

\begin{figure}[bht]
\includegraphics[width=1\columnwidth]{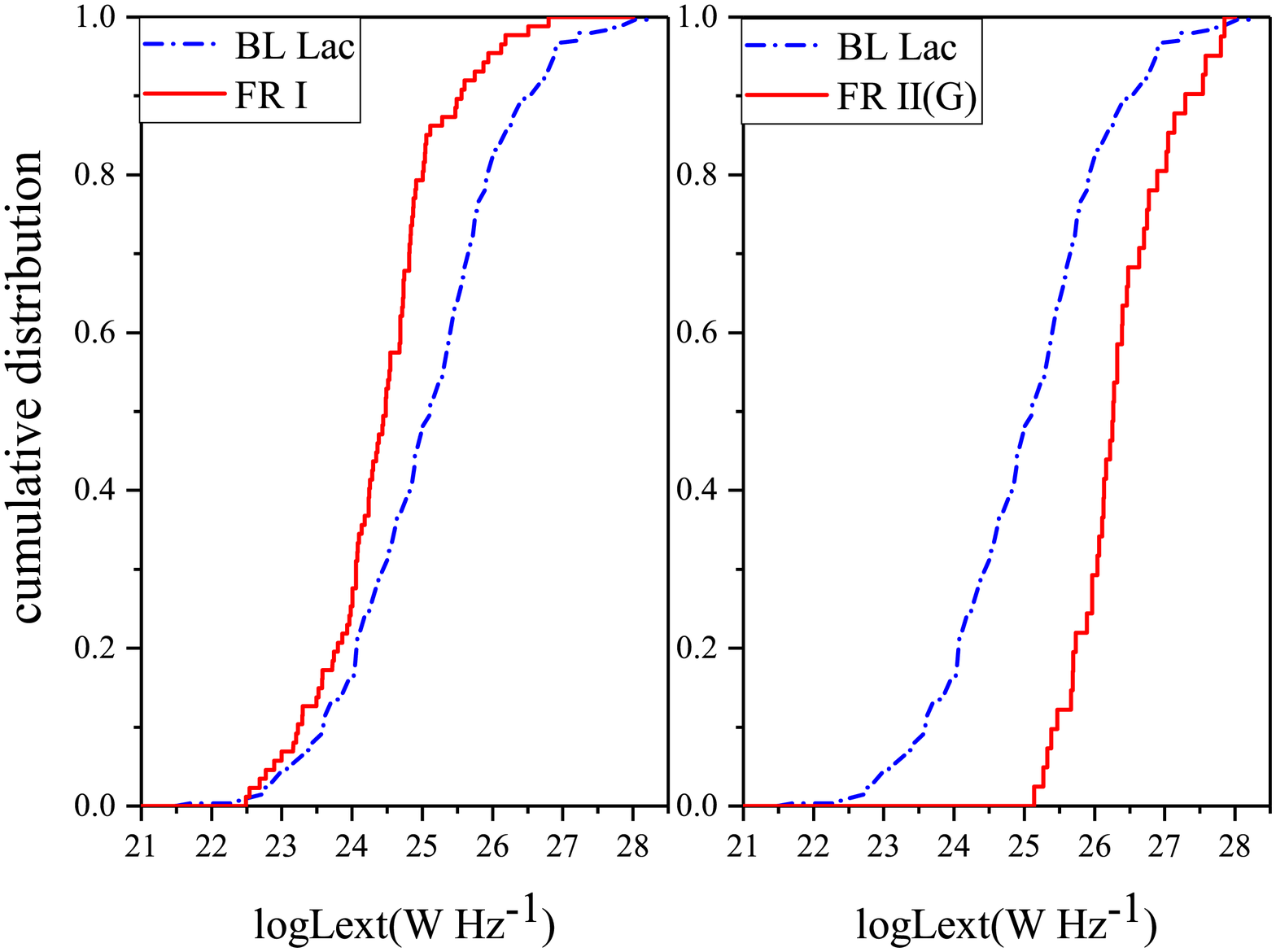}
\caption{The left panel is the cumulative distributions for extended luminosity of BL Lacs and FR Is, in which the  solid line represents FR Is and broken-dotted line represents BL Lacs. The right panel is the cumulative distributions for extended luminosity of BL Lacs and FR II(G)s, in which the solid line represents FR II(G)s and broken-dotted line represents BL Lacs.}
\label{fig.B-Fext}
\end{figure}

As for BL Lacs and FR Is, the extended luminosity distribution of BL Lacs ($\log L_{\rm{ext}}$ = 21.56 $--$ 28.19) is wider than that of FR Is ($\log L_{\rm{ext}}$ = 22.49 $--$ 26.80) and BL Lacs have a higher average  extended luminosity than that of FR Is.  There is a marginal difference for low luminosity of BL Lacs and FR Is, but clear difference for high luminosity as shown in the left panel of Fig. \ref{fig.B-Fext}, the probability values for the K-S test and WRS test are both $<10^{-4}$, in which the hypothesis for WRS test as did \citet{lioda18}  is to determine whether the two independent samples are from the same distribution and for K-S  test is to discriminate between two statistical distributions, their p value threshold is taken as 0.05. 

The extended luminosities for BL Lacs and FR II(G)s are also discussed. The extended luminosities for FR II(G)s are almost the high luminosity ($\log L_{\rm{ext}}$ = 25.14 $--$ 27.85), so the cumulative distribution for extended luminosity of BL Lacs renders discrepant result from that of FR II(G)s with both p $<10^{-4}$ for K-S test and WRS test as shown in right panel of Fig. \ref{fig.B-Fext}.

\begin{figure}[bht]
\subfigure[]{
\label{fig:fig5a}
\includegraphics[width=0.75\columnwidth]{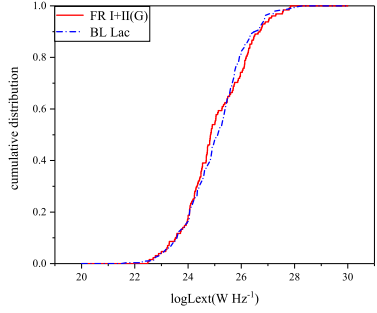}}
\subfigure[]{
\label{fig:fig5b}
\includegraphics[width=0.75\columnwidth]{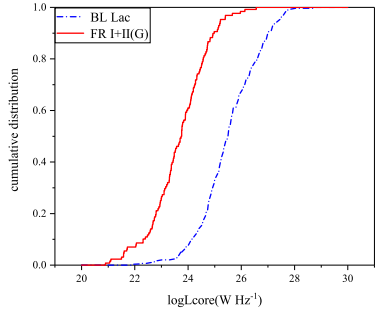}}
\caption{The cumulative distributions of BL Lacs and FR I/II(G)s: \subref{fig:fig5a} for extended luminosity. \subref{fig:fig5b} for core luminosity. The  solid line represents the FR I/II(G)s, and the broken-dotted line represents the BL Lacs. }
\label{fig5}
\end{figure}

It is proposed that the FR Is and FR II(G)s are the parent population of BL Lacs as we mentioned above \citep{xie93,owen96,fan97}.  If it is the case, one can expect that, for the extended luminosity, the distribution of FR I/II(G)s and that of BL Lacs should be from the same parent distribution. Now, we will investigate those distributions using WRS test and K-S test. For the FR I/II(G)s, the total extended luminosity is in a range of $\log L_{\rm{ext}}$ = 22.49 $--$ 27.85 with an average value 25.04, and their cumulative distribution for extended luminosity is close to that of BL Lacs, in which the corresponding cumulative distributions for extended luminosity of BL Lacs and FR I/II(G)s are shown in Fig. \ref{fig5}\subref{fig:fig5a}. When the WRS test and K-S test are adopted to the  distribution of extended luminosity of BL Lacs and that of FR I/II(G)s, it is found that the  probabilities for the both extended luminosities  to be from the same parent distribution are $p_{\rm{WRS}}=0.779$ and  $p_{\rm{K-S}}=0.326$,  implying that the null hypothesis cannot be reject, suggesting that the extended luminosity  distribution for BL Lacs and that of FR I/II(G)s are from the same parent distribution, which indicates that the BL Lacs and FR I/II(G)s are unified. 

We also compared the core luminosity for BL Lacs and FR I/II(G)s. The total core luminosity for FR I/II(G)s is spanning from $\log L_{\rm{core}}$ = 20.90 to 26.56 with an average of 23.67, while the  core luminosity for BL Lac is from $\log L_{\rm{core}}$ = 21.96 to 28.70 with an average of  25.53. The core luminosity distribution for FR I/II(G)s is significantly different from that of BL Lacs as presented in Fig. \ref{fig5}\subref{fig:fig5b}. Both the WRS test and K-S test give p $<10^{-4}$, showing that the core luminosity of BL Lacs and that of FR I/II(G)s are from different distributions. The discrepancy for core luminosity between BL Lacs and FR I/II(G)s is due to the strong beaming effect in BL Lacs. 

\section{Estimation of Doppler Factor }
\subsection{Methodology}

When the jet direction of a BL Lac is close to our line of sight,  it causes a strong beaming effect for the core emission, 
$S^{\rm{ob}}_{\rm{core}} = \delta^q S^{\rm{in}}_{\rm{core}}$,  where $ S^{\rm{ob}}_{\rm{core}}$ and $S^{\rm{in}}_{\rm{core}}$  are the observed core flux density and the intrinsic core flux density of BL Lacs, 
$q = 2 + \alpha_c$  or $3 + \alpha_c$  from different jet structure \citep{sch79}, and  $\alpha_c$ is the radio core spectral index of BL Lacs ($S_{\rm{core, \nu}} \propto \nu^{-\alpha_c}$). 
For luminosities, one can get $L_{\rm{BL,core}}^{\rm {ob}} = \delta ^q L_{\rm{BL,core}}^{\rm {in}}$, in which the $L_{\rm{BL,core}}^{\rm {ob}}$ and $L_{\rm{BL,core}}^{\rm {in}}$ are the observed core luminosity and intrinsic core luminosity of BL Lacs.  

As we mentioned in \S 2.3, BL Lacs and FR I/II(G)s are unified.  In this sense,  one can expect that the intrinsic luminosity of BL Lacs and that of FR I/II(G)s should be from the same parent population,  and that the intrinsic luminosity of BL Lacs and that of FR I/II(G)s should follow the same correlation. Both core luminosity and extended luminosity depend on redshift, and a significant linear correlation ($\log L_{\rm{ext}}=1.18\log L_{\rm{core}}-4.8$) between core and extended luminosity for radio galaxies  is shown in \citet{koll96}. So, we assume that there is a linear correlation between core luminosity ($L_{\rm{FR,core}}$) and extended luminosity ($L_{\rm{FR,ext}}$) for FR I/II(G)s
\begin{equation}
\log L_{\rm{FR,core}} =  k \log L_{\rm{FR,ext}} + b. 
\end{equation}
where k and b are the slope and intercept  for FR I/II(G)s.  It is clear that the intrinsic luminosity in FR I/II(G)s are almost the same as the observed luminosity since the beaming effect is very weak. While for BL Lacs, one has observed core luminosity  enhanced by Doppler factor, $L_{\rm{BL,core}}^{\rm {ob}} = \delta^q L_{\rm{BL,core}}^{\rm {in}}$, and extended luminosity shows as intrinsic luminosity. From unification of BL Lacs and FR I/II(G)s, the intrinsic core luminosity ($L_{\rm{BL,core}}^{\rm {in}}$) and extended luminosity ($L_{\rm{BL,ext}}$) for BL Lacs should  follow the same correlation as do FR I/II(G)s, namely $\log L_{\rm{BL,core}}^{\rm {in}} = k \log L_{\rm{BL,ext}} + b$. 
So, we can get the Doppler factor, $\delta$, for BL Lacs,
\begin{equation}
\log \delta=(\log L_{\rm{BL,c o r e}}^{\rm{o b}}- k  \log L_{\rm{BL,ext}}-b) / q
\end{equation}

\subsection{Luminosity Correlation}
\begin{figure}[bht]
\subfigure[]{
\label{fig:fig6a}
\includegraphics[width=0.8\columnwidth]{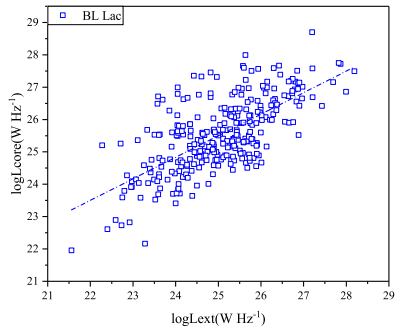}}
\quad
\subfigure[]{
\label{fig:fig6b}
\includegraphics[width=0.8\columnwidth]{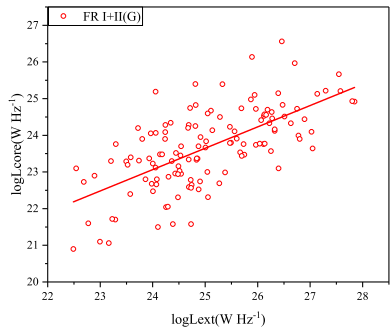}}
\caption{Distributions of core and extended luminosity: \subref{fig:fig6a} for BL Lacs, the broken-dotted line represents the best linear regression between core and extended luminosity for BL Lacs. \subref{fig:fig6b} for FR I/II(G)s, the solid line represents the  best linear regression  between core and extended luminosity for FR I/II(G)s. }
\label{core-ex}
\end{figure}

The relation between the core and extended luminosity is shown in Figure \ref{core-ex} \subref{fig:fig6a} and \subref{fig:fig6b} for the BL Lacs and FR I/II(G)s, respectively.

The correlation coefficient r = 0.678 and p $< 10^{-4}$ are obtained between the core and extended luminosities of BL Lacs. When a linear regression is adopted to BL Lacs, we found that.
\begin{equation}
\log L^{\rm{ob}}_{\rm{BL,core}}=(0.67\pm0.04) \log L_{\rm{BL,ext}}+8.86\pm1.05 \label{bl}
\end{equation}

As for FR I/II(G)s, the r = 0.656 and p $< 10^{-4}$ are obtained between the core and extended luminosities. When a linear regression is performed to the core and extended luminosities of FR I/II(G)s, we obtained following correlation. 
\begin{equation}
\log L_{\rm{FR,core}}=(0.58\pm0.06) \log L_{\rm{FR,ext}}+9.08\pm1.51 \label{fr}
\end{equation}
The best fitting results are shown in Figure \ref{core-ex} \subref{fig:fig6a}, \subref{fig:fig6b}.  It is shown that there are moderate correlations between  core and extended luminosities for both BL Lacs and FR I/II(G)s respectively. 
Since the differences in the slope and intercept between BL Lacs and FR I/II(G)s are smaller than three times the fitting error, the slope and intercept of  BL Lacs are consistent to those of FR I/II(G)s within the fitting errors.

\subsection{Doppler Factor}
In the unified model, FR I/II(G)s are the parent population of BL Lacs. It means that BL Lacs are the FR radio galaxies with the jets pointing to the observers and boosted. As mentioned in \S 3.1, one can get an expression for a Doppler factor based on the slope k (0.58), intercept b (9.08) for the linear correlation of FR I/II(G)s,
\begin{equation}
\log \delta=(\log L_{\rm{BL,c o r e}}^{\rm{o b}}-0.58\log L_{\rm{BL,ext}}-9.08) / q
\end{equation}
When radio core spectral index $\alpha_c = 0.0$ \citep{dona01,abdo10,fan16} is adopted, 
Doppler factors are obtained in a range from $\delta=0.62$ to 113.08 with an average of 15.03 for $q=2$, and from $\delta=0.73$ to 23.38 with an average of 5.43 for $q=3$, which are listed in Col. 2-3 in Table \ref{tab2}. The Doppler factors from different literatures for BL Lacs are also presented in Table \ref{tab2}.

\begin{figure}[bht]
\includegraphics[width=1\columnwidth]{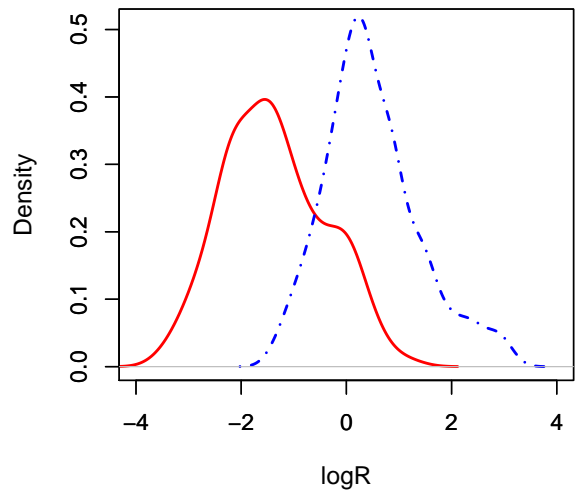}
\caption{The density distributions of logarithm of core-dominance parameter of BL Lacs and FR I/II(G)s. The broken-dotted line represents BL Lacs and the solid line represents FR I/II(G)s.}
\label{ratio}
\end{figure}

In addition, based on the unification of BL Lacs and FR I/II(G)s, we propose another method about the core to extended emission ratio of FR I/II(G)s to estimate Doppler factor. The core to extended emission ratio is also called as core-dominance parameter (R), using the expression, $R=S_{\rm{core}} / S_{\rm{ext}}$, or $R=L_{\rm{core}} / L_{\rm{ext}}$.  If the core emission is strongly boosted by beaming effect, a close relation between the core-dominance parameter (R) and Doppler factor ($\delta$) should be expected, which have been proposed   by \citet{ghi93}: $R=f\delta^q$,  where the $f=S^{\rm{in}}_{\rm{core}}/S_{\rm{ext}}$, $q=2+\alpha_c$ or $3+\alpha_c$ (see above). When an unification of BL Lacs, FR Is and FR II(G)s is considered, the intrinsic core to extended emission ratio ($f_{\rm{BL}}$) of BL Lacs should behave as the same as those ($R_{\rm{FR}}$) of  FR I/II(G)s, then a Doppler factor can be estimated from following correlation. 
\begin{equation}
\delta^q=R_{\rm{BL}}/R_{\rm{FR}}
\end{equation}
where $R_{\rm{FR}}$ is the core-dominance parameter ($R_{\rm{FR}}=L_{\rm{FR,core}}/L_{\rm{FR,ext}}$) for the FR I/II(G)s, and $R_{\rm{BL}}$ is the observed core-dominance parameter ($R_{\rm{BL}}=L_{\rm{BL,core}}^{\rm{ob}}/L_{\rm{BL,ext}}$) for the BL Lacs. The density distributions of logarithm of core-dominance parameter of BL Lacs and FR I/II(G)s are shown in Figure \ref{ratio}. For the present FR I/II(G)s sample, they shows a peak at log $R_{\rm{FR}}$ = -1.48 in Figure \ref{ratio}. When we used this value to estimate the Doppler factors for the case of $q= 2+\alpha_c$ and $3+\alpha_c$, the Doppler factor values are listed in Col. 4 and Col. 5 in Table \ref{tab2}.

\setlength{\tabcolsep}{3pt}
\begin{table}[bht]
  \tbl{$\delta$ from methods and different literatures for BL Lacs}{%
  \begin{tabular}{lcccccccc}
      \hline
      &q=2\footnotemark[$\dag$]&q=3\footnotemark[$\dag$]&q=2\footnotemark[$\ddag$]&q=3\footnotemark[$\ddag$]&G93&H09&L18&Z20 \\
      (1)&(2)&(3)&(4)&(5)&(6)&(7)&(8)&(9)\\   
      \hline
       Min&  0.62   & 0.73 &1.15&1.10 &0.01  &1.1   & 0.22  &0.35     \\
       Max&  113.08   &  23.38 &201.01&34.32  &14.3  &24   & 60.36  &53.57     \\
       Medium&  8.25  & 4.08 &8.24&5.93  &2.1  &6.3   & 9.78    &7.09     \\
       Mean& 15.03   & 5.43&18.23&4.08   &3.85  &7.9  & 13.03 &10.32     \\
      \hline
    \end{tabular}}\label{tab2}
\begin{tabnote}
\footnotemark[$\dag$] The Doppler factor estimation by the core to extended  fitting/regression method.

\footnotemark[$\ddag$] The Doppler factor estimation by the core to extended flux ratio method. 
\tabnote{Note:  Col. 1 is parameters of samples, Col. 2 Doppler factor at q = 2 by fitting/regression method, Col. 3 Doppler factor at q = 3 by fitting/regression method, Col. 4 Doppler factor at q = 2 by ratio method, Col. 5 Doppler factor at q = 3 by ratio method, Col. 6 Doppler factor by \citet{ghi93}, Col. 7 Doppler factor by \citet{hova09}, Col. 8 Doppler factor by \citet{lioda18}, Col. 9 Doppler factor by \citet{zhang20}.}
\end{tabnote}
\end{table}

From different assumptions and methods, it can be found that some Doppler factors for BL Lacs are smaller than unity, which may be caused by the systematic error or the limitation of the methods. For examples, there are 8 BL Lacs in \citet{ghi93}, 5 BL Lacs in \citet{lioda18}, and 3 BL Lacs in \citet{zhang20} with Doppler factor $\delta<1$.  \citet{lioda18} adopted a definite intrinsic brightness temperature to estimate the Doppler factor for BL Lacs. If some BL Lacs was in a low state when it was observed,  it is possible to obtain a low observed brightness temperature, then to derive a Doppler factor $\delta<1$ from the limitation of the methods. In our sample, the source 1440+356 with Doppler factor $\delta<1$ in fitting/regression method may also be due to our systematic error or the limitation of our method that the unification of BL Lacs and FR I/II(G)s.

\section{Discussions}
BL Lacs show special observation properties, such as variability, high and variable polarization, high luminosity,  high energetic $\gamma$-ray emissions, or superluminal motion etc. Their special observation properties are due to the beaming effect.  When their jets are perpendicular to the line of sight, they are radio galaxies, and FR Is and FR II(G)s are proposed to be the parent population of BL Lacs \citep{xie93,owen96,fan97}.

From the available extended luminosities of BL Lacs and FR I/II(G)s, we can see that the probabilities from WRS test and K-S test render discrepant results, p = 0.779 for WRS test and p = 0.326 for K-S test. These results suggest, for the case of the extended luminosity, that BL Lacs are unified with FR I/II(G)s.  But there is also a different probability by the WRS test and K-S test, which may be due to the difference in extended luminosity distributions of BL Lacs and FR I/II(G)s. The extended luminosity distribution of BL Lacs is similar to a normal distribution, but that of FR I/II(G)s is marginally different to normal distribution, in particular from $\log L_{\rm{ext}}= 25$ to 26 shown in Fig. {\ref{normal}. The asymptotic relative efficiency of the WRS test compared to the {\it t-test} is 0.955 for normal distributions,
\begin{figure}[bht]
\includegraphics[width=1\columnwidth]{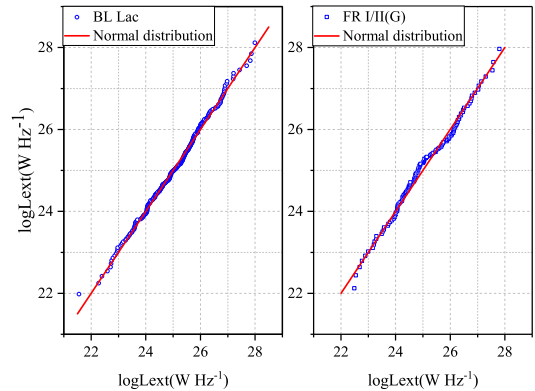}
\caption{The left panel is the comparison between the extended luminosity distribution of BL Lacs and normal distribution, in which the  solid line represents normal distribution and the circular represents BL Lacs. The right panel is the comparison between the extended luminosity distribution of FR I/II(G)s and normal distribution, in which the solid line represents normal distribution and the square represents FR I/II(G)s.}
\label{normal}
\end{figure}
indicating that it can be effectively used for both normal and nonnormal situations \citep{fei12}, while K-S test is not. So the marginal difference of extended luminosity distributions between BL Lacs and FR I/II(G)s is represented in probability that the p value (0.779) for WRS test is higher than that (0.326) of K-S test.

However, the averaged value of the core luminosity of BL Lacs is higher than that of FR I/II(G)s, both WRS test and K-S test suggest that the probabilities for the distribution of the logarithm of the core luminosities of FR I/II(G)s and that of BL Lacs to come from the same parent distribution are very low, they are both $p < 10^{-4}$. The clear difference is from the fact that the core luminosities of BL Lacs are strongly beamed.

As mentioned in \S 1, many methods were proposed to estimate the Doppler factors  \citep{ghi93,mat93,lah99,fan13,chen18,lioda18,zhang20}. In the present work, based on the unified scheme of BL Lacs and FR I/II(G)s, the Doppler factor of BL Lacs is estimated. For BL Lacs objects, the radio band spectrum is flat. For example, \citet{fan06} obtained the radio average value of radio spectral index to be $0.235$ for X-ray selected BL Lacs, and 0.044  for radio selected BL Lacs. \citet{pei16} studied a samples of 1335 blazars and showed the average radio spectral index, $<\alpha_{\rm{radio}}>$ = $0.02\pm0.31$ for {\it Fermi}-detected BL Lacs, $0.34\pm0.37$ for non-{\it Fermi}-detected BL Lacs. $-0.5\leq \alpha \leq 0.5 $ was adopted for BL Lacs in \citet{yuan14}. $\alpha_c$ = 0.0 was also adopted for BL Lacs by \citet{dona01,abdo10}, and \citet{fan16}. We adopted the radio core spectral index $\alpha_c$ = 0.0 for the radio spectral index. By comparing the Doppler factor between $\alpha_c=0$ and $\alpha_c=\pm0.5$, we can find that the Doppler factor for the case $\alpha_c=\pm0.5$ could be several times different from that of $\alpha_c=0$ for $q=2$. But for $q=3$, there is marginal difference between $\alpha_c=\pm0.5$ and $\alpha_c=0$ to estimate the Doppler factor. The Doppler factor corresponding $\alpha_c=0$ or $\pm0.5$ are listed in Col. 10-15 in Table 1.

The coefficients of regression lines have some degrees of errors as shown in Equations (\ref{bl}) and (\ref{fr}). When we estimated the Doppler factor using the fitting/regression method, the coefficient errors of regression lines do have a great influence, even on the order of magnitude, on the Doppler factor value, but the average fitting/regression should be representative of the true Doppler factor values.

Following the case of the spherical jet (q = 3)\citep{ghi93,xie93,hova09,lioda18} and the radio core spectral index as 0 \citep{dona01,abdo10,fan16}, we can also compare our Doppler factor estimation results ($\delta_{\rm{TW}}$) with those  ($\delta_{\rm{G}}$ by \citet{ghi93}, $\delta_{\rm{H}}$ by \citet{hova09}, $\delta_{\rm{L}}$ by \citet{lioda18}, $\delta_{\rm{Z}}$ by \citet{zhang20}) from the literatures for the common sources. There are 29 sources in common with \citet{ghi93}, we performed a linear regression and obtained $\delta_{\rm{TW}} = (0.78\pm0.13) \delta_{\rm{G}} + 4.84\pm0.75$, with a Spearman's rank correlation coefficient of $r = 0.541$ and a chance probability of $p = 2.4\times10^{-3}$, see the upper-left panel in Figure \ref{fig9}. 
We also performed the regression for the common sources with \citet{hova09} $\delta_{\rm{TW}} = (0.70\pm0.13) \delta_{\rm{H}} + 4.39\pm1.10$, \citet{lioda18} $\delta_{\rm{TW}} = (0.17\pm0.05) \delta_{\rm{L}} + 4.88\pm0.67$ and \citet{zhang20} $\delta_{\rm{TW}} = (0.41\pm0.09) \delta_{\rm{Z}} + 3.86\pm1.27$. The Spearman's rank correlation coefficients and chance probabilities are
$r = 0.656$ and $p = 3\times10^{-3}$  for 18 sources with  \citet{hova09}; 
$r = 0.537$ and $p = 3\times 10^{-4} $ for 41 sources with \citet{lioda18}, and 
$r = 0.537$ and $p = 0.016$ for 20 sources with \citet{zhang20}. The best fitting results are all shown in Figure \ref{fig9}.
The correlation coefficients for our Doppler factor estimation results with other literatures for common sources are both larger than 0.5 with probabilities $<0.05$, indicating that our Doppler factor estimation by fitting/regression method is correlated with other samples.
\begin{figure}[bht]
\includegraphics[width=1\columnwidth]{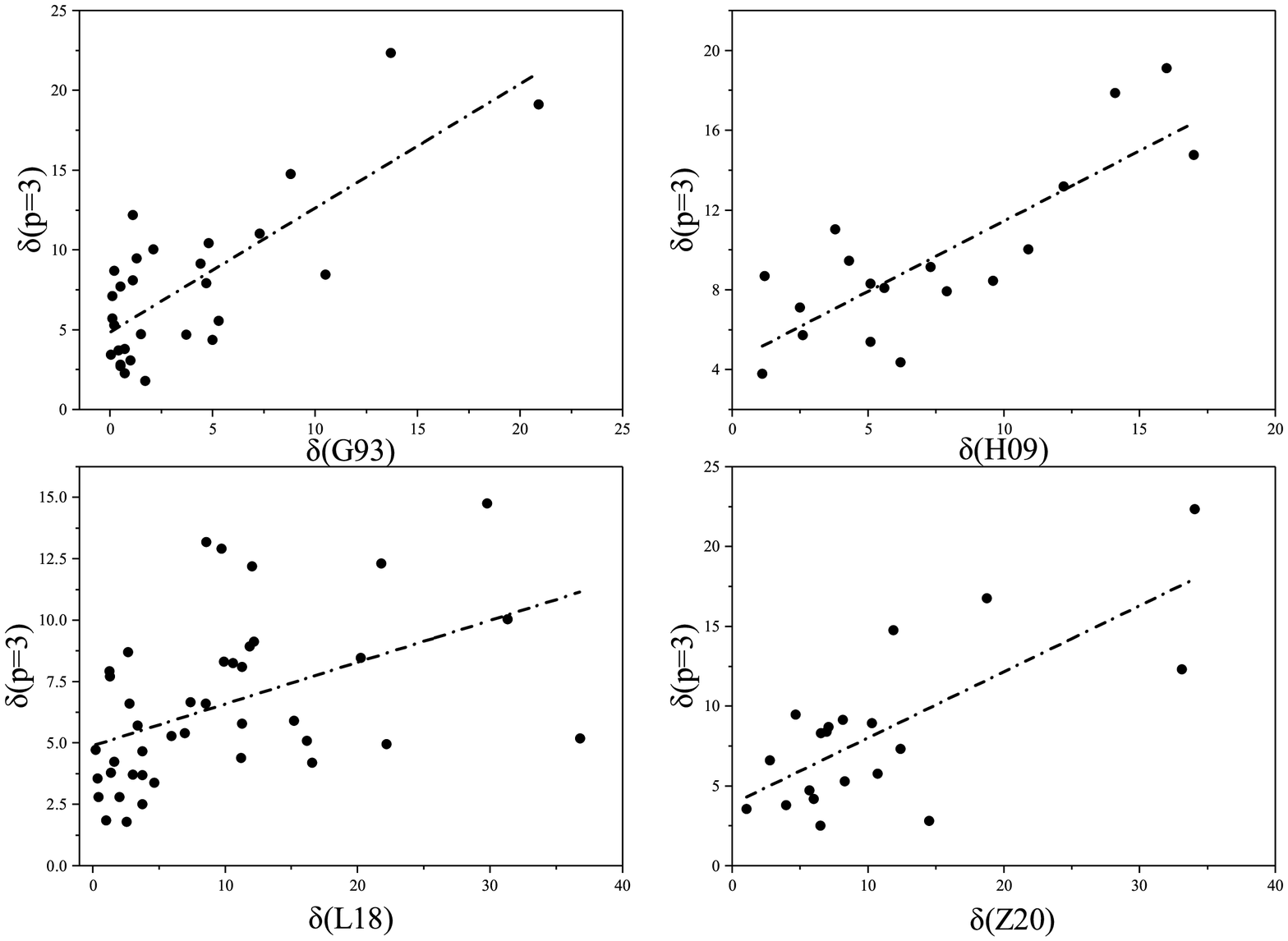}
\caption{Plot of the correlation for common sources between our Doppler factor ($\delta_{\rm{TW}}$) by the fitting/regression method and those from \citet{ghi93}, $\delta_{\rm{G}}$, \citet{hova09}, $\delta_{\rm{H}}$, \citet{lioda18}, $\delta_{\rm{L}}$, \citet{zhang20}, $\delta_{\rm{Z}}$. } 
\label{fig9}
\end{figure}

We also compare the correlations for common sources with other Doppler factor ($\delta_{G}$, $\delta_{H}$, $\delta_{L}$, $\delta_{Z}$) for
Doppler factor (q=3) estimation of the core to extended flux ratio method 
as do the core to extended luminosity fitting/regression method. The Spearman's rank correlation coefficients between the ratio method and those samples are $r=0.347$ and $p=0.07$ for \citet{ghi93}, $r=0.411$, $p=0.09$ for \citet{hova09},   $r=0.182$, $p=0.254$ for \citet{lioda18}, and $r=0.513$, $p=0.022$ for \citet{zhang20}. 
A detailed comparison between the ratio method and fitting/regression method is listed in Table \ref{tab3}. From Table \ref{tab3}, a comparison of the distributions of the core to extended flux ratio is straightforward, but the correlation coefficients of this ratio method are less relevant and less convincing than those of fitting/regression method as mentioned above.

\setlength{\tabcolsep}{11pt}
\begin{table}[bht]
  \tbl{A comparison between the fitting/regression method and the ratio method.}{%
  \begin{tabular}{ccccc}
      \hline
    samples  &$r_f$\footnotemark[$\dag$]&$p_f$\footnotemark[$\dag$]&$r_r$\footnotemark[$\ddag$]&$p_r$\footnotemark[$\ddag$] \\
      (1)&(2)&(3)&(4)&(5)\\   
      \hline
       G93&  0.541   & $2.4\times10^{-3}$   &0.347  &0.07     \\
       H09&  0.656   & $3\times10^{-3}$  &0.411  &0.09       \\
       L18&  0.537  & $3\times10^{-4}$  &0.182 &0.254   \\
       Z20& 0.537 & 0.016   &0.513  & 0.022     \\
      \hline
    \end{tabular}}\label{tab3}
\begin{tabnote}
\footnotemark[$\dag$] The correlation coefficients and probabilities for fitting/regression method.

\footnotemark[$\ddag$] The correlation coefficients and probabilities for ratio method. 
\tabnote{Note:  Col. 1 are samples, Col. 2 the correlation coefficients for fitting/regression method, Col. 3 the probabilities for fitting/regression method, Col. 4 the correlation coefficients for ratio method, Col. 5 the probabilities for ratio method.}
\end{tabnote}
\end{table}

\section{Conclusions}

In this work, we compiled the core and extend flux densities for a sample of BL Lacs, FR Is and FR II(G)s, and calculated their corresponding luminosities. We used WRS test and K-S test to analyze the cumulative distribution for extended luminosity, $L_{\rm{ext}}$, of BL Lacs and that of FR I/II(G)s, and found that the probabilities for the both to be from the same distribution are  $p_{\rm{WRS}}=0.779$ and $p_{\rm{K-S}}=0.326$. Based on the unification of BL Lacs and FR I/II(G)s, we proposed a core to extended luminosities fitting/regression method and a ratio method of core to extended emissions to estimate the Doppler factor for BL Lacs, and compared our results with those in the literatures. 
Our conclusions are as follows:

1. From the extended radio luminosities, BL Lacs are unified with FR I and FR II(G) radio galaxies, which confirmed the results by \citet{xie93}, \citet{owen96}, \citet{fan97}.

2. Our Doppler factors from the fitting/regression method is correlated with those by \citet{ghi93}, \citet{hova09}, \citet{lioda18}, and \citet{zhang20}.

3. The Doppler factors of BL Lacs estimated by the fitting/regression method is in a range of from  $\delta$ = 0.73 to  $\delta$ = 23.38 for the case of q = 3, $\alpha_c=0$.
\begin{ack}
The work is supported by the National Natural Science Foundation of China (NSFC U2031201, NSFC 11733001, NSFC U1938110, NSFC U1531245), Natural Science Foundation of Guangdong Province (2019B030302001), Guangzhou University (NO YM2020001, No 2019GDJC-D18), and supports for Astrophysics Key Subjects of Guangdong Province and Guangzhou City. We thank the anonymous referee for the comments that made us improve our manuscript.
\end{ack}

\appendix 
\section*{The complete Table 1 sample.}

\setcounter{table}{0}
\setlength{\tabcolsep}{2pt}
\begin{longtable}{lcccccccccccccc}
\caption{The core and extended fluxes or luminosities of whole samples and the Doppler factor of BL Lacs.}
\hline
Source&class&$z$ & $S_{\rm{core}}$ & $S_{\rm{ext}}$&Ref&$logL_{\rm{core}}$&$logL_{\rm{ext}}$&Ref&$\delta_2$ &$\delta_2		$&$\delta_2$&$\delta_3$&$\delta_3$&$\delta_3$ \\
Name& & &(mJy)&(mJy)& &(W\ Hz$^{-1})$&(W\ Hz$^{-1})$  & &($\alpha_c=0$ &0.5&-0.5)&($\alpha_c=0$&0.5&-0.5) \\
(1)&(2)&\,(3)&(4)&\,\,(5)&(6)&(7)&(8)&(9)&(10)&(11)&(12)&(13)&(14)&(15)\\
\hline
\endfirsthead
\hline
Source&class&$z$ & $S_{\rm{core}}$ & $S_{\rm{ext}}$&Ref&$logL_{\rm{core}}$&$logL_{\rm{ext}}$&Ref&$\delta_2$ &$\delta_2		$&$\delta_2$&$\delta_3$&$\delta_3$&$\delta_3$ \\
Name& & &(mJy)&(mJy)& &(W\ Hz$^{-1})$&(W\ Hz$^{-1})$  & &($\alpha_c=0$ &0.5&-0.5)&($\alpha_c=0$&0.5&-0.5) \\
(1)&(2)&\,(3)&(4)&\,\,(5)&(6)&(7)&(8)&(9)&(10)&(11)&(12)&(13)&(14)&(15)\\
\hline
\endhead 
\hline
\endfoot 
0003+003	&	B	&	1.037	&	480	&	206	&	P19	&	27.17	&	26.81	&		&	18.74	&	10.43	&	49.79 	&	7.06	&	5.34	&	10.43 	\\
0003-066	&	B	&	0.347	&	1850	&	1504.47	&	P19	&	26.73	&	26.64	&		&	12.55	&	7.57	&	29.16 	&	5.40	&	4.24	&	7.57 	\\
0007+124	&	I	&	0.156	&	4	&	751.02	&	F11	&	23.47	&	25.75	&		&		&		&		&		&		&		\\
0007+472	&	B	&	0.280	&	67	&	22	&	P19	&	25.17	&	24.69	&		&	7.69	&	5.11	&	15.17 	&	3.90	&	3.21	&	5.11 	\\
0011+1853	&	B	&	0.473	&	140	&	104	&	P19	&	25.89	&	25.76	&		&	8.58	&	5.58	&	17.57 	&	4.19	&	3.42	&	5.58 	\\
0013+790	&	II(G)	&	0.840	&	4.4	&	1028.6	&	F11	&	25.21	&	27.58	&		&		&		&		&		&		&		\\
0021+055	&	B	&	2.050	&	28	&	53	&	P19	&	26.56	&	26.83	&		&	9.04	&	5.82	&	18.84 	&	4.34	&	3.52	&	5.82 	\\
0029-271	&	B	&	0.333	&	11	&	105	&	P19	&	24.67	&	25.65	&		&	2.27	&	1.93	&	2.98 	&	1.73	&	1.60	&	1.93 	\\
0032+595	&	B	&	0.086	&	44	&	5	&	P19	&	23.92	&	22.97	&		&	5.71	&	4.03	&	10.20 	&	3.19	&	2.71	&	4.03 	\\
0033+156	&	B	&	1.162	&	125	&	28	&	P19	&	26.96	&	26.31	&		&	20.43	&	11.18	&	55.86 	&	7.47	&	5.61	&	11.18 	\\
0038+328	&	II(G)	&	0.482	&	0.47	&	1199.53	&	NED	&	23.64	&	27.05	&		&		&		&		&		&		&		\\
0039+398	&	I	&	0.109	&	1	&	225	&	P19	&	22.52	&	24.88	&		&		&		&		&		&		&		\\
0043+008	&	B	&	2.149	&	2	&	2	&	P19	&	25.27	&	25.27	&		&	5.86	&	4.11	&	10.55 	&	3.25	&	2.75	&	4.11 	\\
0044+193	&	B	&	0.181	&	7	&	17	&	P19	&	23.71	&	24.10	&		&	2.13	&	1.83	&	2.74 	&	1.65	&	1.54	&	1.83 	\\
0048-09	&	B	&	0.634	&	887	&	108.45	&	F11	&	26.98	&	26.07	&		&	24.62	&	12.97	&	71.62 	&	8.46	&	6.24	&	12.97 	\\
0052+251	&	B	&	0.154	&	1	&	1	&	P19	&	22.73	&	22.73	&		&	1.71	&	1.54	&	2.05 	&	1.43	&	1.36	&	1.54 	\\
0053+260	&	I	&	0.195	&		&		&		&	23.00	&	25.04	&	F11	&		&		&		&		&		&		\\
0055-01	&	I	&	0.045	&	93	&	2043.47	&	M93	&	23.67	&	25.01	&		&		&		&		&		&		&		\\
0057+026	&	B	&	0.599	&	99	&	16	&	P19	&	25.83	&	25.04	&		&	13.02	&	7.80	&	30.65 	&	5.54	&	4.34	&	7.80 	\\
0057+3021	&	I	&	0.017	&		&		&		&	23.40	&	23.58	&	B11	&		&		&		&		&		&		\\
0059+581	&	B	&	0.644	&	1570	&	7	&	P19	&	27.32	&	24.96	&		&	75.51	&	31.80	&	319.16 	&	17.86	&	11.83	&	31.80 	\\
0104+32	&	I	&	0.016	&		&		&		&	22.04	&	24.25	&	Z95	&		&		&		&		&		&		\\
0106+130	&	II(G)	&	0.060	&	38	&	5140.8	&	F11	&	23.53	&	25.67	&		&		&		&		&		&		&		\\
0106+729	&	II(G)	&	0.181	&		&		&		&	23.76	&	26.11	&	F11	&		&		&		&		&		&		\\
0107+32224	&	I	&	0.017	&		&		&		&	22.80	&	23.86	&	B11	&		&		&		&		&		&		\\
0109+224	&	B	&	0.265	&	330	&	5.44	&	F11	&	25.82	&	24.04	&		&	25.12	&	13.18	&	73.56 	&	8.58	&	6.31	&	13.18 	\\
0109+492	&	II(G)	&	0.395	&	3.64	&	734.59	&	NED	&	24.33	&	26.63	&		&		&		&		&		&		&		\\
0115-261	&	I	&	0.053	&	18	&	5	&	P19	&	23.10	&	22.54	&		&		&		&		&		&		&		\\
0118-272	&	B	&	0.559	&	1137	&	63	&	P19	&	26.97	&	25.71	&		&	30.74	&	15.49	&	96.28 	&	9.81	&	7.08	&	15.49 	\\
0120+340	&	B	&	0.272	&	31	&	2	&	P19	&	24.81	&	23.62	&		&	10.35	&	6.48	&	22.55 	&	4.75	&	3.80	&	6.48 	\\
0121+318	&	B	&	0.654	&	82	&	89	&	P19	&	26.08	&	26.11	&		&	8.42	&	5.50	&	17.14 	&	4.14	&	3.38	&	5.50 	\\
0122+090	&	B	&	0.339	&	1	&	1	&	P19	&	23.52	&	23.52	&		&	2.51	&	2.09	&	3.41 	&	1.85	&	1.69	&	2.09 	\\
0123+3315	&	I	&	0.016	&		&		&		&	20.90	&	22.49	&	B11	&		&		&		&		&		&		\\
0125+287	&	II(G)	&	0.437	&	185	&	105	&	NED	&	26.14	&	25.89	&		&		&		&		&		&		&		\\
0125-0120	&	I	&	0.018	&		&		&		&	22.80	&	24.08	&	B11	&		&		&		&		&		&		\\
0138-097	&	B	&	0.733	&	696	&	504	&	P19	&	26.93	&	26.79	&		&	14.33	&	8.42	&	34.82 	&	5.90	&	4.58	&	8.42 	\\
0140+219B	&	B	&	0.599	&	7	&	13	&	P19	&	24.89	&	25.16	&		&	4.07	&	3.07	&	6.49 	&	2.55	&	2.23	&	3.07 	\\
0145+138	&	B	&	0.125	&	2	&	1	&	P19	&	22.89	&	22.59	&		&	2.27	&	1.92	&	2.98 	&	1.72	&	1.60	&	1.92 	\\
0154+286	&	II(G)	&	0.735	&	5.3	&	536.7	&	NED	&	25.13	&	27.14	&		&		&		&		&		&		&		\\
0156+0537	&	I	&	0.019	&		&		&		&	21.70	&	23.29	&	B11	&		&		&		&		&		&		\\
0158+003	&	B	&	0.299	&	9	&	1	&	P19	&	24.37	&	23.41	&		&	7.14	&	4.82	&	13.74 	&	3.71	&	3.07	&	4.82 	\\
0159+002	&	B	&	0.163	&	7	&	12	&	P19	&	23.70	&	23.94	&		&	2.34	&	1.98	&	3.12 	&	1.77	&	1.63	&	1.98 	\\
0200-0011	&	B	&	0.366	&	30	&	88	&	P19	&	25.04	&	25.51	&		&	3.82	&	2.92	&	5.98 	&	2.45	&	2.15	&	2.92 	\\
0204+29	&	I	&	0.110	&	171	&	158	&	P19	&	24.75	&	24.71	&		&		&		&		&		&		&		\\
0208+352	&	B	&	0.318	&	5	&	1	&	P19	&	24.31	&	23.61	&		&	5.85	&	4.11	&	10.54 	&	3.25	&	2.74	&	4.11 	\\
0208-512	&	B	&	0.999	&	233	&	3169	&	P19	&	26.86	&	27.99	&		&	5.91	&	4.14	&	10.69 	&	3.27	&	2.76	&	4.14 	\\
0212+364	&	B	&	0.490	&	82	&	1	&	P19	&	25.54	&	23.63	&		&	23.91	&	12.67	&	68.86 	&	8.30	&	6.13	&	12.67 	\\
0213-132	&	I	&	0.147	&	99	&	98	&	P19	&	24.82	&	24.82	&		&		&		&		&		&		&		\\
0214+083	&	B	&	1.400	&	296	&	166	&	P19	&	27.12	&	26.87	&		&	16.94	&	9.62	&	43.51 	&	6.60	&	5.04	&	9.62 	\\
0219+042	&	I	&	0.022	&	161	&	130	&	P19	&	23.30	&	23.21	&		&		&		&		&		&		&		\\
0219+428	&	B	&	0.444	&	814	&	510.84	&	F11	&	26.75	&	26.55	&		&	13.67	&	8.10	&	32.68 	&	5.72	&	4.46	&	8.10 	\\
0220+427	&	I	&	0.021	&		&		&		&	22.59	&	24.69	&	F11	&		&		&		&		&		&		\\
0221+27	&	II(G)	&	0.310	&	20	&	884	&	F11	&	24.83	&	26.48	&		&		&		&		&		&		&		\\
0227+020	&	B	&	0.457	&	9	&	18	&	P19	&	24.76	&	25.06	&		&	3.74	&	2.87	&	5.80 	&	2.41	&	2.12	&	2.87 	\\
0230+344	&	B	&	0.458	&	123	&	56	&	P19	&	25.89	&	25.55	&		&	9.92	&	6.27	&	21.31 	&	4.62	&	3.71	&	6.27 	\\
0232+202	&	B	&	0.599	&	43	&	7	&	P19	&	25.68	&	24.89	&		&	12.06	&	7.33	&	27.65 	&	5.26	&	4.15	&	7.33 	\\
0232+264	&	B	&	0.599	&	554	&	1	&	P19	&	26.79	&	24.05	&		&	76.10	&	32.00	&	322.50 	&	17.96	&	11.89	&	32.00 	\\
0240+1656	&	B	&	0.599	&	40	&	7	&	P19	&	25.65	&	24.89	&		&	11.63	&	7.12	&	26.35 	&	5.13	&	4.06	&	7.12 	\\
0245+269A	&	B	&	0.599	&	28	&	1	&	P19	&	25.49	&	24.05	&		&	17.11	&	9.70	&	44.09 	&	6.64	&	5.07	&	9.70 	\\
0247-027	&	I	&	0.087	&		&		&		&	23.45	&	24.53	&	Z95	&		&		&		&		&		&		\\
0255+05	&	I	&	0.024	&		&		&		&	22.65	&	24.74	&	Z95	&		&		&		&		&		&		\\
0257+342	&	B	&	0.245	&	9	&	2	&	P19	&	24.15	&	23.49	&		&	5.25	&	3.77	&	9.12 	&	3.02	&	2.58	&	3.77 	\\
0300+162	&	I	&	0.033	&	8	&	1188.74	&	F11	&	22.31	&	24.48	&		&		&		&		&		&		&		\\
0300+470	&	B	&	0.475	&	1566	&	62.16	&	F11	&	26.95	&	25.55	&		&	33.63	&	16.65	&	108.56 	&	10.42	&	7.45	&	16.65 	\\
0301-243	&	B	&	0.260	&	228	&	132	&	F11	&	25.67	&	25.43	&		&	8.30	&	5.44	&	16.81 	&	4.10	&	3.35	&	5.44 	\\
03020-037	&	I	&	0.005	&	28	&	102	&	P19	&	23.37	&	23.93	&		&		&		&		&		&		&		\\
0305+03	&	I	&	0.029	&	964	&	2436	&	DM14	&	24.28	&	24.68	&		&		&		&		&		&		&		\\
0307+169	&	II(G)	&	0.256	&	6.04	&	963.14	&	F11	&	24.12	&	26.32	&		&		&		&		&		&		&		\\
0308+0406	&	I	&	0.029	&		&		&		&	24.20	&	24.67	&	B11	&		&		&		&		&		&		\\
0308+104	&	B	&	0.599	&	886	&	1	&	P19	&	26.99	&	24.05	&		&	96.24	&	38.61	&	441.05 	&	21.00	&	13.59	&	38.61 	\\
0314+063	&	B	&	0.599	&	27	&	2	&	P19	&	25.48	&	24.35	&		&	13.74	&	8.14	&	32.91 	&	5.74	&	4.47	&	8.14 	\\
0315+41	&	I	&	0.026	&	40	&	3490	&	P19	&	22.78	&	24.72	&		&		&		&		&		&		&		\\
0317+185	&	B	&	0.190	&	17	&	15	&	P19	&	24.22	&	24.17	&		&	3.65	&	2.82	&	5.62 	&	2.37	&	2.10	&	2.82 	\\
0319+4130	&	I	&	0.018	&		&		&		&	25.40	&	24.81	&	B11	&		&		&		&		&		&		\\
0320-37	&	I	&	0.006	&	51	&	71949	&	DM14	&	21.58	&	24.73	&		&		&		&		&		&		&		\\
0329+704	&	B	&	0.599	&	37	&	12	&	P19	&	25.53	&	25.04	&		&	9.19	&	5.90	&	19.25 	&	4.39	&	3.55	&	5.90 	\\
0331-001	&	I	&	0.139	&		&		&		&	24.53	&	26.12	&	Z95	&		&		&		&		&		&		\\
0350-371	&	B	&	0.165	&	14	&	2.9	&	P19	&	24.04	&	23.35	&		&	5.08	&	3.67	&	8.74 	&	2.96	&	2.53	&	3.67 	\\
0356+102	&	I	&	0.030	&	9	&	4961	&	P19	&	22.31	&	25.05	&		&		&		&		&		&		&		\\
036-019	&	B	&	0.850	&	14	&	2	&	P19	&	25.52	&	24.67	&		&	11.59	&	7.10	&	26.23 	&	5.12	&	4.06	&	7.10 	\\
0402-014	&	B	&	0.920	&	876	&	102	&	P19	&	27.39	&	26.45	&		&	30.35	&	15.34	&	94.68 	&	9.73	&	7.03	&	15.34 	\\
0410+110	&	II(G)	&	0.306	&		&		&		&	25.15	&	26.39	&	F11	&		&		&		&		&		&		\\
0414+009	&	B	&	0.287	&		&		&		&	24.76	&	25.30	&	F11	&	3.18	&	2.53	&	4.68 	&	2.16	&	1.94	&	2.53 	\\
0414+378	&	I	&	0.049	&	252	&	230	&	P19	&	24.28	&	24.24	&		&		&		&		&		&		&		\\
0419+194	&	B	&	0.512	&	8	&	1	&	P19	&	24.77	&	23.87	&		&	8.39	&	5.48	&	17.05 	&	4.13	&	3.37	&	5.48 	\\
0422+004	&	B	&	0.310	&	1465	&	1.75	&	F11	&	26.61	&	23.69	&		&	78.75	&	32.88	&	337.52 	&	18.37	&	12.12	&	32.88 	\\
0430+05	&	I	&	0.033	&		&		&		&	25.19	&	24.05	&	Z95	&		&		&		&		&		&		\\
0433+29	&	II(G)	&	0.218	&	108.9	&	13063.54	&	F11	&	25.22	&	27.29	&		&		&		&		&		&		&		\\
0449-175	&	I	&	0.031	&		&		&		&	21.58	&	24.38	&	Z95	&		&		&		&		&		&		\\
0453+22	&	II(G)	&	0.214	&	4.1	&	937.86	&	F11	&	23.77	&	26.13	&		&		&		&		&		&		&		\\
0454+844	&	B	&	1.340	&	392	&	1346.15	&	F11	&	27.16	&	27.69	&		&	10.17	&	6.40	&	22.03 	&	4.69	&	3.76	&	6.40 	\\
0502+675	&	B	&	0.314	&	17	&	4	&	P19	&	24.68	&	24.05	&		&	6.69	&	4.58	&	12.61 	&	3.55	&	2.96	&	4.58 	\\
0521-365	&	B	&	0.055	&	3124	&	5213.24	&	F11	&	25.41	&	25.63	&		&	5.39	&	3.85	&	9.46 	&	3.08	&	2.62	&	3.85 	\\
0545-199	&	I	&	0.053	&		&		&		&	23.23	&	24.00	&	Z95	&		&		&		&		&		&		\\
0548-322	&	B	&	0.069	&	80	&	92.73	&	F11	&	24.01	&	24.07	&		&	3.04	&	2.43	&	4.40 	&	2.10	&	1.89	&	2.43 	\\
0605+48	&	II(G)	&	0.277	&	0.5	&	996.1	&	F11	&	23.10	&	26.40	&		&		&		&		&		&		&		\\
0607+710	&	B	&	0.267	&	12	&	11	&	P19	&	24.41	&	24.38	&		&	3.96	&	3.01	&	6.27 	&	2.50	&	2.20	&	3.01 	\\
0620-52	&	I	&	0.051	&	260	&	946.39	&	M93	&	24.25	&	24.81	&		&		&		&		&		&		&		\\
0625-35	&	I	&	0.055	&	600	&	1650	&	DM14	&	24.67	&	25.11	&		&		&		&		&		&		&		\\
0634-205	&	I	&	0.055	&		&		&		&	22.74	&	24.90	&	Z95	&		&		&		&		&		&		\\
0647+250	&	B	&	0.203	&	42	&	36	&	P19	&	24.68	&	24.61	&		&	4.59	&	3.38	&	7.62 	&	2.76	&	2.39	&	3.38 	\\
0648-165	&	B	&	0.599	&	2120	&	4	&	P19	&	27.33	&	24.60	&		&	97.49	&	39.01	&	448.70 	&	21.18	&	13.69	&	39.01 	\\
0651+542	&	II(G)	&	0.238	&	2	&	963.2	&	F11	&	23.57	&	26.25	&		&		&		&		&		&		&		\\
0702+749	&	II(G)	&	0.292	&	9.64	&	631.90	&	F11	&	24.46	&	26.28	&		&		&		&		&		&		&		\\
0704+384	&	B	&	0.579	&	61	&	260	&	P19	&	25.93	&	26.56	&		&	5.28	&	3.79	&	9.20 	&	3.03	&	2.59	&	3.79 	\\
0708+7413	&	B	&	0.371	&	65	&	151	&	P19	&	25.32	&	25.68	&		&	4.69	&	3.44	&	7.84 	&	2.80	&	2.42	&	3.44 	\\
0716+714	&	B	&	0.300	&	2460	&	88.04	&	K05	&	26.77	&	25.33	&		&	31.75	&	15.90	&	100.55 	&	10.03	&	7.21	&	15.90 	\\
0722+30	&	I	&	0.019	&		&		&		&	22.90	&	22.89	&	Z95	&		&		&		&		&		&		\\
0723-008	&	B	&	0.128	&		&		&		&	24.89	&	25.95	&	F11	&	2.40	&	2.01	&	3.21 	&	1.79	&	1.65	&	2.01 	\\
0734+805	&	II(G)	&	0.118	&	7	&	1271.2	&	F11	&	23.44	&	25.70	&		&		&		&		&		&		&		\\
0735+178	&	B	&	0.424	&	1919	&	59.39	&	K10	&	26.98	&	25.47	&		&	36.63	&	17.83	&	121.67 	&	11.03	&	7.83	&	17.83 	\\
0737+746	&	B	&	0.315	&	21	&	1	&	P19	&	24.76	&	23.43	&		&	11.02	&	6.82	&	24.53 	&	4.95	&	3.94	&	6.82 	\\
0738+5451	&	B	&	0.720	&	279	&	1	&	P19	&	26.63	&	24.18	&		&	57.64	&	25.62	&	222.65 	&	14.92	&	10.14	&	25.62 	\\
0742+333	&	B	&	0.611	&	124	&	1	&	P19	&	26.13	&	24.04	&		&	35.92	&	17.55	&	118.53 	&	10.89	&	7.74	&	17.55 	\\
0743+7458	&	B	&	0.607	&	22	&	11	&	P19	&	25.41	&	25.11	&		&	7.65	&	5.09	&	15.08 	&	3.88	&	3.20	&	5.09 	\\
0744+55	&	I	&	0.036	&	105	&	1596	&	P19	&	23.84	&	25.02	&		&		&		&		&		&		&		\\
0749+540	&	B	&	0.200	&	737	&	133.98	&	F11	&	25.89	&	25.15	&		&	12.90	&	7.74	&	30.26 	&	5.50	&	4.31	&	7.74 	\\
0754+100	&	B	&	0.266	&	1087	&	57.73	&	F11	&	26.35	&	25.07	&		&	23.04	&	12.30	&	65.56 	&	8.10	&	6.01	&	12.30 	\\
0759+508	&	B	&	0.054	&	8	&	10	&	P19	&	22.82	&	22.92	&		&	1.68	&	1.51	&	2.00 	&	1.41	&	1.34	&	1.51 	\\
0800+244	&	I	&	0.040	&	4	&	135	&	P19	&	22.48	&	24.00	&		&		&		&		&		&		&		\\
0806+505	&	B	&	1.207	&	12	&	11	&	P19	&	25.83	&	25.79	&		&	7.87	&	5.21	&	15.65 	&	3.96	&	3.25	&	5.21 	\\
0806+524	&	B	&	0.138	&	66	&	70.8	&	P19	&	24.52	&	24.55	&		&	3.98	&	3.02	&	6.31 	&	2.51	&	2.20	&	3.02 	\\
0808+019	&	B	&	1.148	&	424	&	11.16	&	F11	&	27.22	&	25.64	&		&	43.14	&	20.32	&	151.29 	&	12.30	&	8.59	&	20.32 	\\
0810+5619	&	B	&	0.510	&	49	&	1	&	P19	&	25.61	&	23.92	&		&	21.25	&	11.53	&	58.86 	&	7.67	&	5.73	&	11.53 	\\
0812+578	&	B	&	0.054	&	46	&	18	&	P19	&	23.58	&	23.17	&		&	3.39	&	2.66	&	5.10 	&	2.26	&	2.01	&	2.66 	\\
0812+6217	&	B	&	0.599	&	20	&	18	&	P19	&	25.43	&	25.39	&		&	6.52	&	4.48	&	12.18 	&	3.49	&	2.92	&	4.48 	\\
0818-128	&	B	&	0.074	&	270	&	540	&	F11	&	24.59	&	24.89	&		&	3.44	&	2.69	&	5.20 	&	2.28	&	2.03	&	2.69 	\\
0819+525	&	B	&	0.599	&	23	&	20	&	P19	&	25.41	&	25.35	&		&	6.50	&	4.47	&	12.14 	&	3.48	&	2.92	&	4.47 	\\
0820+225	&	B	&	0.951	&		&		&		&	26.42	&	27.43	&	F11	&	5.19	&	3.73	&	8.99 	&	3.00	&	2.56	&	3.73 	\\
0824+294	&	II(G)	&	0.458	&	116.1	&	636.9	&	F11	&	25.97	&	26.71	&		&		&		&		&		&		&		\\
0824+4204	&	B	&	0.223	&	7	&	40	&	P19	&	24.01	&	24.77	&		&	1.92	&	1.68	&	2.38 	&	1.54	&	1.45	&	1.68 	\\
0826+180	&	B	&	0.089	&		&		&		&	23.96	&	24.50	&	F11	&	2.16	&	1.85	&	2.80 	&	1.67	&	1.55	&	1.85 	\\
0828+493	&	B	&	0.548	&		&		&		&	25.90	&	26.66	&	F11	&	4.77	&	3.49	&	8.03 	&	2.83	&	2.44	&	3.49 	\\
0829+046	&	B	&	0.174	&	643	&	103.90	&	F11	&	25.69	&	24.90	&		&	12.13	&	7.37	&	27.88 	&	5.28	&	4.16	&	7.37 	\\
0837-12	&	B	&	0.198	&	160	&	672	&	F11	&	25.28	&	25.90	&		&	3.87	&	2.95	&	6.07 	&	2.46	&	2.17	&	2.95 	\\
0847+548	&	B	&	0.367	&	6	&	47	&	P19	&	24.45	&	25.34	&		&	2.16	&	1.85	&	2.79 	&	1.67	&	1.55	&	1.85 	\\
0850+443	&	B	&	0.382	&	31	&	46	&	P19	&	25.17	&	25.34	&		&	4.96	&	3.60	&	8.45 	&	2.91	&	2.50	&	3.60 	\\
0850+625	&	B	&	0.267	&	224	&	1	&	P19	&	25.68	&	23.33	&		&	34.26	&	16.90	&	111.26 	&	10.55	&	7.53	&	16.90 	\\
0851+203	&	B	&	0.306	&	1719	&	7	&	P19	&	26.67	&	24.28	&		&	56.72	&	25.29	&	217.92 	&	14.76	&	10.05	&	25.29 	\\
0855+082	&	B	&	0.455	&	32	&	82	&	P19	&	25.39	&	25.80	&		&	4.72	&	3.46	&	7.91 	&	2.81	&	2.43	&	3.46 	\\
0905-097	&	B	&	0.053	&		&		&		&	23.69	&	23.59	&	F11	&	2.91	&	2.35	&	4.15 	&	2.04	&	1.84	&	2.35 	\\
0906+041	&	B	&	3.200	&	78	&	33	&	P19	&	27.58	&	27.21	&		&	22.94	&	12.26	&	65.18 	&	8.07	&	5.99	&	12.26 	\\
0908+445	&	B	&	0.298	&	31	&	126	&	P19	&	24.97	&	25.58	&		&	3.36	&	2.64	&	5.04 	&	2.24	&	2.00	&	2.64 	\\
0912+297	&	B	&	0.101	&	222	&	79.23	&	F11	&	24.78	&	24.33	&		&	6.22	&	4.32	&	11.44 	&	3.38	&	2.84	&	4.32 	\\
0915+32	&	I	&	0.062	&		&		&		&	23.12	&	24.06	&	Z95	&		&		&		&		&		&		\\
0915-118	&	I	&	0.065	&		&		&		&	24.51	&	26.51	&	Z95	&		&		&		&		&		&		\\
0917+45	&	II(G)	&	0.174	&	44	&	1873	&	F11	&	24.64	&	26.27	&		&		&		&		&		&		&		\\
0922+3625	&	B	&	1.015	&	167	&	1	&	P19	&	26.77	&	24.54	&		&	53.16	&	24.01	&	199.87 	&	14.14	&	9.68	&	24.01 	\\
0923+750	&	B	&	0.638	&	5	&	52	&	P19	&	24.81	&	25.83	&		&	2.37	&	2.00	&	3.16 	&	1.78	&	1.64	&	2.00 	\\
0925+504	&	B	&	0.370	&	462	&	615.03	&	F11	&	26.16	&	26.29	&		&	8.29	&	5.43	&	16.77 	&	4.10	&	3.35	&	5.43 	\\
0926+2550	&	B	&	0.539	&	113	&	8	&	P19	&	26.04	&	24.89	&		&	18.25	&	10.21	&	48.04 	&	6.93	&	5.26	&	10.21 	\\
0927+352	&	B	&	0.435	&	394	&	47	&	F11	&	26.33	&	25.41	&		&	18.12	&	10.15	&	47.59 	&	6.90	&	5.24	&	10.15 	\\
0927+500	&	B	&	0.187	&	15	&	7	&	P19	&	24.14	&	23.81	&		&	4.22	&	3.17	&	6.83 	&	2.61	&	2.28	&	3.17 	\\
0944+734	&	I	&	0.058	&	131	&	129	&	P19	&	24.06	&	24.06	&		&		&		&		&		&		&		\\
0945+222	&	B	&	0.716	&	48	&	79	&	P19	&	26.00	&	26.21	&		&	7.18	&	4.84	&	13.86 	&	3.72	&	3.09	&	4.84 	\\
0945+664	&	B	&	0.850	&	1407	&	33	&	P19	&	27.57	&	25.94	&		&	52.64	&	23.82	&	197.26 	&	14.05	&	9.63	&	23.82 	\\
0946+003	&	B	&	0.585	&	108	&	3	&	P19	&	26.07	&	24.51	&		&	24.31	&	12.84	&	70.41 	&	8.39	&	6.19	&	12.84 	\\
0950+495	&	B	&	0.380	&	4	&	1	&	P19	&	24.22	&	23.61	&		&	5.25	&	3.77	&	9.12 	&	3.02	&	2.58	&	3.77 	\\
0951+216	&	B	&	0.296	&	33	&	303	&	P19	&	24.94	&	25.90	&		&	2.62	&	2.16	&	3.61 	&	1.90	&	1.73	&	2.16 	\\
0952+226A	&	B	&	1.211	&	17	&	117	&	P19	&	25.92	&	26.76	&		&	4.57	&	3.37	&	7.59 	&	2.75	&	2.38	&	3.37 	\\
0954+65	&	B	&	0.368	&	637	&	746.53	&	F11	&	26.29	&	26.36	&		&	9.15	&	5.88	&	19.14 	&	4.37	&	3.54	&	5.88 	\\
0958+290	&	II(G)	&	0.185	&	34.46	&	1238.36	&	F11	&	24.57	&	26.13	&		&		&		&		&		&		&		\\
0958+294	&	B	&	0.558	&	142	&	30	&	P19	&	26.12	&	25.45	&		&	13.84	&	8.18	&	33.23 	&	5.76	&	4.49	&	8.18 	\\
0958+426A	&	B	&	0.664	&	23	&	37	&	P19	&	25.51	&	25.71	&		&	5.71	&	4.03	&	10.21 	&	3.20	&	2.71	&	4.03 	\\
1003+328	&	B	&	1.026	&	58	&	173	&	P19	&	26.43	&	26.90	&		&	7.44	&	4.98	&	14.53 	&	3.81	&	3.15	&	4.98 	\\
1003+351	&	II(G)	&	0.101	&	900	&	772	&	F11	&	25.40	&	25.33	&		&		&		&		&		&		&		\\
1009+427	&	B	&	0.365	&	29	&	17	&	P19	&	25.09	&	24.86	&		&	6.26	&	4.34	&	11.53 	&	3.40	&	2.85	&	4.34 	\\
1011+446	&	B	&	0.796	&	7	&	15	&	P19	&	25.33	&	25.66	&		&	4.83	&	3.53	&	8.17 	&	2.86	&	2.46	&	3.53 	\\
1011+496	&	B	&	0.212	&		&		&		&	24.91	&	24.98	&	F11	&	4.69	&	3.44	&	7.84 	&	2.80	&	2.42	&	3.44 	\\
1015+383	&	B	&	0.387	&	16	&	101	&	P19	&	24.92	&	25.72	&		&	2.89	&	2.34	&	4.12 	&	2.03	&	1.83	&	2.34 	\\
1020+493	&	B	&	0.390	&	12	&	52	&	P19	&	24.73	&	25.37	&		&	2.95	&	2.37	&	4.22 	&	2.06	&	1.85	&	2.37 	\\
1027+555A	&	B	&	0.435	&	7	&	73	&	P19	&	24.72	&	25.74	&		&	2.27	&	1.93	&	2.98 	&	1.73	&	1.60	&	1.93 	\\
1028+511	&	B	&	0.360	&	23	&	11	&	P19	&	24.93	&	24.61	&		&	6.14	&	4.27	&	11.23 	&	3.35	&	2.82	&	4.27 	\\
1030+585	&	II(G)	&	0.428	&	1.38	&	837.37	&	NED	&	23.99	&	26.78	&		&		&		&		&		&		&		\\
1034+5727	&	B	&	0.830	&	89	&	37	&	P19	&	26.13	&	25.75	&		&	11.44	&	7.03	&	25.78 	&	5.08	&	4.03	&	7.03 	\\
1040+31	&	I	&	0.036	&		&		&		&	23.48	&	24.14	&	Z95	&		&		&		&		&		&		\\
1044+549	&	B	&	0.540	&	4	&	2	&	P19	&	24.57	&	24.27	&		&	5.08	&	3.67	&	8.74 	&	2.96	&	2.53	&	3.67 	\\
1055+0519	&	B	&	0.890	&	179	&	25	&	P19	&	26.62	&	25.76	&		&	19.85	&	10.92	&	53.75 	&	7.33	&	5.52	&	10.92 	\\
1055+567	&	B	&	0.143	&	178	&	69	&	P19	&	24.98	&	24.57	&		&	6.68	&	4.57	&	12.58 	&	3.55	&	2.96	&	4.57 	\\
1101+384	&	B	&	0.030	&	520	&	156.65	&	F11	&	24.09	&	23.57	&		&	4.68	&	3.44	&	7.83 	&	2.80	&	2.42	&	3.44 	\\
1101+411	&	B	&	0.035	&	13	&	8	&	P19	&	22.61	&	22.40	&		&	1.86	&	1.64	&	2.28 	&	1.51	&	1.42	&	1.64 	\\
1106+244	&	B	&	0.482	&	18	&	1	&	P19	&	25.16	&	23.90	&		&	12.81	&	7.69	&	29.96 	&	5.47	&	4.29	&	7.69 	\\
1116+227	&	B	&	0.422	&	104	&	34	&	P19	&	25.67	&	25.18	&		&	9.81	&	6.21	&	20.99 	&	4.58	&	3.69	&	6.21 	\\
1118+424	&	B	&	0.124	&	19	&	11	&	P19	&	23.87	&	23.64	&		&	3.49	&	2.72	&	5.28 	&	2.30	&	2.04	&	2.72 	\\
1122+39	&	I	&	0.007	&		&		&		&	21.06	&	23.16	&	Z95	&		&		&		&		&		&		\\
1133+704	&	B	&	0.045	&	131	&	220.48	&	F11	&	23.83	&	24.06	&		&	2.50	&	2.08	&	3.40 	&	1.84	&	1.69	&	2.08 	\\
1142+198	&	I	&	0.022	&	250	&	2107.56	&	F11	&	23.52	&	24.44	&		&		&		&		&		&		&		\\
1144+352	&	B	&	0.063	&	537	&	126	&	F11	&	24.74	&	24.12	&		&	6.90	&	4.69	&	13.14 	&	3.63	&	3.02	&	4.69 	\\
1144-379	&	B	&	1.048	&	2182	&	18	&	P19	&	27.66	&	25.58	&		&	74.60	&	31.49	&	314.06 	&	17.72	&	11.75	&	31.49 	\\
1145+1936	&	I	&	0.021	&		&		&		&	23.30	&	24.36	&	B11	&		&		&		&		&		&		\\
1147+245	&	B	&	0.200	&	664	&	20	&	F11	&	25.86	&	24.33	&		&	21.41	&	11.60	&	59.46 	&	7.71	&	5.76	&	11.60 	\\
1148+592	&	B	&	0.118	&	95	&	36	&	P19	&	24.55	&	24.13	&		&	5.47	&	3.89	&	9.63 	&	3.10	&	2.64	&	3.89 	\\
1150+449A	&	B	&	0.599	&	9	&	5	&	P19	&	25.00	&	24.74	&		&	6.08	&	4.24	&	11.10 	&	3.33	&	2.81	&	4.24 	\\
1151+6039	&	B	&	1.120	&	75	&	1	&	P19	&	26.69	&	24.82	&		&	40.68	&	19.39	&	139.89 	&	11.83	&	8.31	&	19.39 	\\
1154+435	&	B	&	0.230	&	93	&	13	&	P19	&	25.19	&	24.33	&		&	9.94	&	6.28	&	21.37 	&	4.62	&	3.71	&	6.28 	\\
1202+492	&	B	&	0.452	&	40	&	71	&	P19	&	25.47	&	25.72	&		&	5.45	&	3.88	&	9.58 	&	3.10	&	2.63	&	3.88 	\\
1203+238	&	B	&	0.599	&	8	&	30	&	P19	&	24.95	&	25.52	&		&	3.41	&	2.67	&	5.13 	&	2.27	&	2.02	&	2.67 	\\
1203+645	&	II(G)	&	0.372	&	730	&	579	&	F11	&	26.56	&	26.46	&		&		&		&		&		&		&		\\
1208+322	&	B	&	0.389	&	7	&	149	&	P19	&	24.56	&	25.89	&		&	1.71	&	1.53	&	2.04 	&	1.43	&	1.36	&	1.53 	\\
1210+121	&	B	&	0.369	&	170	&	81.97	&	F11	&	25.91	&	25.59	&		&	9.83	&	6.22	&	21.05 	&	4.59	&	3.69	&	6.22 	\\
1215+303	&	B	&	0.130	&	355	&	157.77	&	F11	&	25.21	&	24.85	&		&	7.17	&	4.83	&	13.82 	&	3.72	&	3.08	&	4.83 	\\
1216+06	&	I	&	0.006	&		&		&		&	22.68	&	23.98	&	Z95	&		&		&		&		&		&		\\
1217+348	&	B	&	0.643	&	258	&	94	&	P19	&	26.43	&	25.99	&		&	13.71	&	8.12	&	32.81 	&	5.73	&	4.46	&	8.12 	\\
1218+285	&	B	&	0.102	&	1118	&	10.52	&	P19	&	25.53	&	23.50	&		&	25.63	&	13.40	&	75.58 	&	8.69	&	6.38	&	13.40 	\\
1218+460	&	B	&	0.525	&	30	&	37	&	P19	&	25.37	&	25.46	&		&	5.78	&	4.07	&	10.36 	&	3.22	&	2.72	&	4.07 	\\
1220+337C	&	B	&	0.599	&	459	&	383	&	P19	&	26.81	&	26.73	&		&	12.98	&	7.78	&	30.52 	&	5.52	&	4.33	&	7.78 	\\
1220+373	&	B	&	0.491	&	15	&	138	&	P19	&	25.17	&	26.13	&		&	2.92	&	2.36	&	4.17 	&	2.04	&	1.84	&	2.36 	\\
1221+245	&	B	&	0.218	&	179	&	25	&	P19	&	25.37	&	24.52	&		&	10.86	&	6.74	&	24.06 	&	4.91	&	3.91	&	6.74 	\\
1221+809	&	B	&	0.369	&	447	&	71	&	F11	&	26.25	&	25.46	&		&	16.04	&	9.21	&	40.45 	&	6.36	&	4.88	&	9.21 	\\
1222+13	&	I	&	0.003	&		&		&		&	21.72	&	23.23	&	F11	&		&		&		&		&		&		\\
1222+488	&	B	&	0.647	&	20	&	18	&	P19	&	25.50	&	25.45	&		&	6.73	&	4.60	&	12.70 	&	3.56	&	2.97	&	4.60 	\\
1227+255	&	B	&	0.135	&	351	&	1	&	P19	&	25.26	&	22.71	&		&	31.76	&	15.90	&	100.57 	&	10.03	&	7.21	&	15.90 	\\
1228+12	&	I	&	0.004	&	3097	&	68469	&	DM14	&	23.34	&	24.69	&		&		&		&		&		&		&		\\
1229.2+6430	&	B	&	0.163	&	42.49	&	3.48	&	F11	&	24.48	&	23.39	&		&	8.25	&	5.41	&	16.67 	&	4.08	&	3.34	&	5.41 	\\
1229+290	&	B	&	0.236	&	60	&	54	&	P19	&	25.00	&	24.95	&		&	5.28	&	3.79	&	9.21 	&	3.03	&	2.59	&	3.79 	\\
1229+405	&	B	&	0.638	&	52	&	57	&	P19	&	25.98	&	26.02	&		&	8.03	&	5.29	&	16.08 	&	4.01	&	3.29	&	5.29 	\\
1235+632	&	B	&	0.297	&	22	&	21	&	P19	&	24.74	&	24.72	&		&	4.60	&	3.39	&	7.64 	&	2.76	&	2.39	&	3.39 	\\
1238+414	&	B	&	0.499	&	10	&	19	&	P19	&	24.84	&	25.11	&		&	3.93	&	2.99	&	6.21 	&	2.49	&	2.19	&	2.99 	\\
1239+069	&	B	&	0.150	&	10	&	1	&	P19	&	23.80	&	22.80	&		&	5.59	&	3.96	&	9.91 	&	3.15	&	2.67	&	3.96 	\\
1243+4402	&	B	&	1.152	&	43	&	5	&	P19	&	26.33	&	25.40	&		&	18.25	&	10.21	&	48.05 	&	6.93	&	5.26	&	10.21 	\\
1246+586	&	B	&	0.847	&	278	&	136	&	F11	&	26.64	&	26.33	&		&	13.94	&	8.23	&	33.56 	&	5.79	&	4.51	&	8.23 	\\
1247+443	&	B	&	1.812	&	12	&	4	&	P19	&	25.90	&	25.42	&		&	10.88	&	6.75	&	24.11 	&	4.91	&	3.91	&	6.75 	\\
1250+532	&	B	&	0.369	&	346	&	50	&	F11	&	26.34	&	25.50	&		&	17.19	&	9.73	&	44.37 	&	6.66	&	5.08	&	9.73 	\\
1251+278	&	II(G)	&	0.086	&		&		&		&	22.99	&	25.38	&	F11	&		&		&		&		&		&		\\
1251-12	&	I	&	0.014	&		&		&		&	22.87	&	24.30	&	Z95	&		&		&		&		&		&		\\
1254+476	&	II(G)	&	0.996	&	1.6	&	1352.96	&	F11	&	24.92	&	27.85	&		&		&		&		&		&		&		\\
1255+244	&	B	&	0.141	&	7	&	1	&	P19	&	23.60	&	22.75	&		&	4.57	&	3.37	&	7.59 	&	2.75	&	2.38	&	3.37 	\\
1259+2757	&	I	&	0.024	&		&		&		&	21.10	&	22.99	&	B11	&		&		&		&		&		&		\\
1259+4112	&	B	&	0.649	&	19	&	14	&	P19	&	25.34	&	25.21	&		&	6.60	&	4.53	&	12.38 	&	3.52	&	2.94	&	4.53 	\\
1302+715	&	B	&	0.599	&	31	&	1	&	P19	&	25.51	&	24.02	&		&	17.80	&	10.01	&	46.47 	&	6.82	&	5.18	&	10.01 	\\
1308+27	&	II(G)	&	0.240	&	2.79	&	472.09	&	F11	&	23.74	&	25.97	&		&		&		&		&		&		&		\\
1309-216	&	B	&	1.491	&	140	&	45	&	F11	&	27.20	&	26.71	&		&	20.66	&	11.27	&	56.69 	&	7.53	&	5.64	&	11.27 	\\
1310+560	&	B	&	0.975	&	235	&	33	&	P19	&	26.99	&	26.14	&		&	23.74	&	12.60	&	68.22 	&	8.26	&	6.11	&	12.60 	\\
1312+240	&	B	&	2.145	&	101	&	31	&	P19	&	27.26	&	26.74	&		&	21.51	&	11.65	&	59.83 	&	7.74	&	5.77	&	11.65 	\\
1318-43	&	I	&	0.011	&		&		&		&	23.48	&	24.18	&	Z95	&		&		&		&		&		&		\\
1319+428	&	II(G)	&	0.079	&		&		&		&	22.69	&	25.27	&	F11	&		&		&		&		&		&		\\
1322+36	&	I	&	0.018	&		&		&		&	23.29	&	23.49	&	Z95	&		&		&		&		&		&		\\
1324+478	&	B	&	0.683	&	15	&	30	&	P19	&	25.29	&	25.59	&		&	4.83	&	3.52	&	8.16 	&	2.86	&	2.46	&	3.52 	\\
1328+506	&	B	&	0.599	&	19	&	3	&	P19	&	25.24	&	24.44	&		&	9.83	&	6.22	&	21.05 	&	4.59	&	3.69	&	6.22 	\\
1338+3851	&	I	&	0.246	&		&		&		&	24.70	&	26.19	&	B11	&		&		&		&		&		&		\\
1339+554	&	B	&	0.207	&	30	&	4	&	P19	&	24.54	&	23.66	&		&	7.37	&	4.94	&	14.33 	&	3.79	&	3.13	&	4.94 	\\
1343-60	&	I	&	0.013	&	2730	&	3850	&	DM14	&	24.09	&	24.24	&		&		&		&		&		&		&		\\
1345+735	&	B	&	0.290	&	17	&	372	&	P19	&	24.67	&	26.01	&		&	1.79	&	1.59	&	2.17 	&	1.47	&	1.39	&	1.59 	\\
1346+26	&	I	&	0.063	&		&		&		&	23.95	&	24.91	&	Z95	&		&		&		&		&		&		\\
1350+4922	&	B	&	0.397	&	63	&	61	&	P19	&	25.38	&	25.36	&		&	6.21	&	4.31	&	11.43 	&	3.38	&	2.84	&	4.31 	\\
1356+393	&	B	&	0.800	&	63	&	31	&	P19	&	26.30	&	25.99	&		&	11.81	&	7.21	&	26.90 	&	5.19	&	4.10	&	7.21 	\\
1400+162	&	B	&	0.244	&	233	&	311.19	&	F11	&	25.62	&	25.75	&		&	6.36	&	4.39	&	11.79 	&	3.43	&	2.88	&	4.39 	\\
1402+042	&	B	&	0.344	&	21	&	12	&	P19	&	24.86	&	24.62	&		&	5.64	&	3.99	&	10.04 	&	3.17	&	2.69	&	3.99 	\\
1404+286	&	B	&	0.077	&	991	&	1909	&	P19	&	25.11	&	25.40	&		&	4.48	&	3.32	&	7.38 	&	2.72	&	2.36	&	3.32 	\\
1407+595	&	B	&	0.496	&	17	&	3	&	P19	&	25.09	&	24.34	&		&	8.88	&	5.74	&	18.39 	&	4.29	&	3.48	&	5.74 	\\
1409+524	&	II(G)	&	0.464	&	10	&	7364	&	NED	&	24.93	&	27.80	&		&		&		&		&		&		&		\\
1413+135	&	B	&	0.247	&	1080	&	4.091	&	K10	&	26.27	&	23.85	&		&	47.86	&	22.08	&	173.76 	&	13.18	&	9.12	&	22.08 	\\
1414+110	&	I	&	0.025	&	56	&	1649.57	&	F11	&	22.96	&	24.43	&		&		&		&		&		&		&		\\
1414+375	&	B	&	0.920	&	41	&	4	&	P19	&	26.17	&	25.16	&		&	17.78	&	10.00	&	46.39 	&	6.81	&	5.18	&	10.00 	\\
1418+546	&	B	&	0.153	&	1058	&	19.82	&	F11	&	25.80	&	24.08	&		&	23.98	&	12.70	&	69.14 	&	8.31	&	6.14	&	12.70 	\\
1420+198	&	II(G)	&	0.270	&	6.2	&	894.32	&	F11	&	24.16	&	26.32	&		&		&		&		&		&		&		\\
1421+582	&	B	&	0.635	&	6	&	89	&	P19	&	24.51	&	25.68	&		&	1.85	&	1.64	&	2.27 	&	1.51	&	1.42	&	1.64 	\\
1422+026	&	I	&	0.037	&	662	&	1868	&	P19	&	24.60	&	25.05	&		&		&		&		&		&		&		\\
1424+240	&	B	&	0.160	&	250	&	60	&	P19	&	25.23	&	24.61	&		&	8.68	&	5.64	&	17.85 	&	4.23	&	3.44	&	5.64 	\\
1426+340	&	B	&	1.553	&	23	&	5	&	P19	&	26.32	&	25.65	&		&	15.08	&	8.76	&	37.26 	&	6.10	&	4.71	&	8.76 	\\
1426+428	&	B	&	0.129	&	19.1	&	2.1	&	P19	&	23.91	&	22.95	&		&	5.75	&	4.05	&	10.30 	&	3.21	&	2.72	&	4.05 	\\
1435+174A	&	B	&	0.599	&	355	&	174	&	P19	&	26.60	&	26.29	&		&	13.65	&	8.09	&	32.61 	&	5.71	&	4.45	&	8.09 	\\
1437+397	&	B	&	0.344	&	38	&	24	&	P19	&	25.06	&	24.86	&		&	6.03	&	4.21	&	10.98 	&	3.31	&	2.79	&	4.21 	\\
1440+122	&	B	&	0.163	&	17.2	&	1.3	&	P19	&	24.09	&	22.97	&		&	6.99	&	4.74	&	13.37 	&	3.66	&	3.04	&	4.74 	\\
1440+356	&	B	&	0.079	&	1	&	13	&	P19	&	22.17	&	23.28	&		&	0.62	&	0.68	&	0.53 	&	0.73	&	0.76	&	0.68 	\\
1441+522	&	II(G)	&	0.722	&	104	&	938.4	&	F11	&	24.74	&	25.69	&		&		&		&		&		&		&		\\
1441+536	&	B	&	2.454	&	17	&	1	&	P19	&	26.77	&	25.54	&		&	27.45	&	14.15	&	82.80 	&	9.10	&	6.64	&	14.15 	\\
1443+634	&	B	&	0.298	&	8	&	7	&	P19	&	24.30	&	24.25	&		&	3.81	&	2.91	&	5.94 	&	2.44	&	2.15	&	2.91 	\\
1446+3620	&	B	&	1.565	&	29	&	1	&	P19	&	26.85	&	25.39	&		&	33.28	&	16.51	&	107.07 	&	10.35	&	7.41	&	16.51 	\\
1447+771	&	II(G)	&	1.132	&	6	&	454	&	F11	&	25.67	&	27.55	&		&		&		&		&		&		&		\\
1448+634	&	I	&	0.042	&		&		&		&	22.57	&	24.73	&	F11	&		&		&		&		&		&		\\
1449+537	&	B	&	0.432	&	6	&	16	&	P19	&	24.57	&	25.00	&		&	3.14	&	2.50	&	4.60 	&	2.14	&	1.92	&	2.50 	\\
1452+516	&	B	&	1.083	&	80	&	53	&	P19	&	26.51	&	26.33	&		&	11.99	&	7.29	&	27.44 	&	5.24	&	4.13	&	7.29 	\\
1454+510	&	B	&	0.599	&	212	&	20	&	P19	&	26.32	&	25.30	&		&	19.27	&	10.67	&	51.68 	&	7.19	&	5.42	&	10.67 	\\
1458+224	&	B	&	0.235	&	60	&	25	&	P19	&	24.88	&	24.50	&		&	6.22	&	4.32	&	11.45 	&	3.38	&	2.84	&	4.32 	\\
1459+551	&	B	&	0.339	&	18	&	56	&	P19	&	24.85	&	25.34	&		&	3.44	&	2.68	&	5.18 	&	2.28	&	2.02	&	2.68 	\\
1501+481	&	B	&	0.345	&	10	&	46	&	P19	&	24.63	&	25.29	&		&	2.75	&	2.25	&	3.85 	&	1.96	&	1.78	&	2.25 	\\
1504+2600	&	I	&	0.054	&		&		&		&	23.70	&	24.87	&	B11	&		&		&		&		&		&		\\
1508+3138	&	B	&	0.672	&	83	&	51	&	P19	&	25.95	&	25.73	&		&	9.33	&	5.97	&	19.65 	&	4.43	&	3.58	&	5.97 	\\
1508+425	&	B	&	0.488	&	19	&	92	&	P19	&	25.22	&	25.91	&		&	3.62	&	2.80	&	5.55 	&	2.36	&	2.08	&	2.80 	\\
1508+561	&	B	&	1.680	&	28	&	16	&	P19	&	26.22	&	25.98	&		&	10.88	&	6.75	&	24.10 	&	4.91	&	3.91	&	6.75 	\\
1508+574	&	B	&	0.817	&	10	&	36	&	P19	&	25.26	&	25.82	&		&	4.02	&	3.04	&	6.39 	&	2.53	&	2.21	&	3.04 	\\
1514+07	&	I	&	0.035	&		&		&		&	24.50	&	25.28	&	Z95	&		&		&		&		&		&		\\
1514+197	&	B	&	1.070	&	255	&	2.62	&	F11	&	26.81	&	24.82	&		&	46.38	&	21.53	&	166.66 	&	12.91	&	8.96	&	21.53 	\\
1514-241	&	B	&	0.049	&	2562	&	3.02	&	F11	&	25.20	&	22.27	&		&	39.97	&	19.11	&	136.64 	&	11.69	&	8.23	&	19.11 	\\
1516+0701	&	I	&	0.034	&		&		&		&	23.90	&	24.23	&	B11	&		&		&		&		&		&		\\
1516+4843	&	B	&	0.576	&	5	&	29	&	P19	&	24.80	&	25.56	&		&	2.79	&	2.27	&	3.93 	&	1.98	&	1.80	&	2.27 	\\
1517+656	&	B	&	0.702	&	19	&	19	&	P19	&	25.33	&	25.33	&		&	6.01	&	4.20	&	10.93 	&	3.31	&	2.79	&	4.20 	\\
1519-273	&	B	&	1.294	&	2290	&	10	&	P19	&	28.00	&	25.64	&		&	105.52	&	41.56	&	498.63 	&	22.33	&	14.33	&	41.56 	\\
1525+29	&	I	&	0.065	&		&		&		&	22.66	&	24.07	&	Z95	&		&		&		&		&		&		\\
1529+5153	&	B	&	0.975	&	21	&	8	&	P19	&	25.70	&	25.28	&		&	9.51	&	6.06	&	20.16 	&	4.49	&	3.62	&	6.06 	\\
1530+190	&	B	&	0.307	&	17	&	11	&	P19	&	24.62	&	24.43	&		&	4.84	&	3.53	&	8.19 	&	2.86	&	2.46	&	3.53 	\\
1532+595	&	B	&	0.599	&	13	&	25	&	P19	&	25.31	&	25.59	&		&	4.93	&	3.58	&	8.39 	&	2.90	&	2.49	&	3.58 	\\
1533+342	&	B	&	0.811	&	33	&	1	&	P19	&	25.86	&	24.34	&		&	21.41	&	11.60	&	59.45 	&	7.71	&	5.76	&	11.60 	\\
1533+535	&	B	&	0.890	&	43	&	1	&	P19	&	25.80	&	24.17	&		&	22.50	&	12.07	&	63.54 	&	7.97	&	5.93	&	12.07 	\\
1534+0147	&	B	&	0.312	&	25	&	9	&	P19	&	24.96	&	24.52	&		&	6.76	&	4.61	&	12.78 	&	3.58	&	2.98	&	4.61 	\\
1534+656	&	B	&	0.539	&	6	&	50	&	P19	&	24.82	&	25.74	&		&	2.54	&	2.11	&	3.46 	&	1.86	&	1.70	&	2.11 	\\
1538+149	&	B	&	0.605	&	1337	&	125.89	&	F11	&	27.17	&	26.15	&		&	29.11	&	14.83	&	89.55 	&	9.46	&	6.86	&	14.83 	\\
1552.1+2020	&	B	&	0.222	&	33.09	&	10.09	&	F11	&	24.68	&	24.16	&		&	6.19	&	4.30	&	11.36 	&	3.37	&	2.83	&	4.30 	\\
1553+113	&	B	&	0.360	&	95	&	29.38	&	P19	&	25.51	&	25.00	&		&	9.21	&	5.91	&	19.31 	&	4.39	&	3.56	&	5.91 	\\
1553-228	&	I	&	0.065	&	126	&	42	&	P19	&	24.20	&	23.72	&		&		&		&		&		&		&		\\
1600+309	&	B	&	1.091	&	20	&	4	&	P19	&	26.07	&	25.37	&		&	13.72	&	8.13	&	32.84 	&	5.73	&	4.47	&	8.13 	\\
1604+1353	&	B	&	0.294	&	32	&	21	&	P19	&	24.89	&	24.70	&		&	5.48	&	3.90	&	9.67 	&	3.11	&	2.64	&	3.90 	\\
1615+351	&	I	&	0.030	&	32	&	68	&	F11	&	23.20	&	23.52	&		&		&		&		&		&		&		\\
1615+412	&	B	&	0.267	&	81	&	43	&	P19	&	25.17	&	24.90	&		&	6.70	&	4.58	&	12.63 	&	3.55	&	2.96	&	4.58 	\\
1618+063	&	B	&	0.435	&	17	&	14	&	P19	&	24.99	&	24.90	&		&	5.40	&	3.85	&	9.47 	&	3.08	&	2.62	&	3.85 	\\
1619+378	&	B	&	1.272	&	155	&	46	&	P19	&	27.18	&	26.65	&		&	20.88	&	11.37	&	57.50 	&	7.58	&	5.68	&	11.37 	\\
1620+103	&	B	&	0.369	&	85	&	55.83	&	F11	&	25.62	&	25.44	&		&	7.81	&	5.18	&	15.50 	&	3.94	&	3.24	&	5.18 	\\
1621+38	&	I	&	0.031	&		&		&		&	23.31	&	23.74	&	Z95	&		&		&		&		&		&		\\
1622+375	&	B	&	0.200	&	14	&	17	&	P19	&	24.20	&	24.28	&		&	3.29	&	2.59	&	4.89 	&	2.21	&	1.97	&	2.59 	\\
1622-253	&	B	&	0.786	&	179	&	25	&	P19	&	26.54	&	25.69	&		&	19.11	&	10.59	&	51.11 	&	7.15	&	5.40	&	10.59 	\\
1625+318	&	B	&	0.732	&	38	&	38	&	P19	&	25.89	&	25.89	&		&	7.89	&	5.22	&	15.71 	&	3.96	&	3.26	&	5.22 	\\
1626+352	&	B	&	0.497	&	14	&	5	&	P19	&	25.02	&	24.58	&		&	7.00	&	4.74	&	13.38 	&	3.66	&	3.04	&	4.74 	\\
1626+396	&	I	&	0.030	&		&		&		&	23.16	&	24.49	&	F11	&		&		&		&		&		&		\\
1626+518	&	I	&	0.055	&	96	&	78	&	P19	&	24.05	&	23.96	&		&		&		&		&		&		&		\\
1629+120	&	B	&	1.795	&	359	&	505	&	P19	&	27.72	&	27.87	&		&	17.30	&	9.78	&	44.74 	&	6.69	&	5.10	&	9.78 	\\
1635+185B	&	B	&	0.599	&	124	&	17	&	P19	&	26.14	&	25.28	&		&	15.83	&	9.11	&	39.75 	&	6.30	&	4.85	&	9.11 	\\
1636+8240	&	I	&	0.025	&	380	&	130	&	DM14	&	23.76	&	23.30	&		&		&		&		&		&		&		\\
1640+396	&	B	&	0.539	&	47	&	5	&	P19	&	25.59	&	24.62	&		&	13.10	&	7.83	&	30.87 	&	5.56	&	4.35	&	7.83 	\\
1643+1715	&	I	&	0.162	&		&		&		&	25.10	&	25.94	&	B11	&		&		&		&		&		&		\\
1651+0459	&	I	&	0.154	&		&		&		&	23.90	&	26.80	&	B11	&		&		&		&		&		&		\\
1652+151	&	B	&	0.290	&	67	&	30	&	P19	&	25.24	&	24.89	&		&	7.26	&	4.88	&	14.05 	&	3.75	&	3.10	&	4.88 	\\
1652+398	&	B	&	0.034	&	1376	&	63.85	&	F11	&	24.59	&	23.26	&		&	10.24	&	6.43	&	22.24 	&	4.72	&	3.78	&	6.43 	\\
1658+471	&	II(G)	&	0.205	&	16.7	&	1296.52	&	F11	&	24.33	&	26.22	&		&		&		&		&		&		&		\\
1659+389	&	B	&	1.113	&	50	&	11	&	P19	&	26.58	&	25.92	&		&	17.06	&	9.67	&	43.93 	&	6.63	&	5.06	&	9.67 	\\
1700+518	&	B	&	0.292	&	2	&	4	&	P19	&	23.72	&	24.02	&		&	2.26	&	1.92	&	2.96 	&	1.72	&	1.59	&	1.92 	\\
1702+298	&	B	&	1.927	&	258	&	306	&	P19	&	27.75	&	27.82	&		&	18.44	&	10.30	&	48.72 	&	6.98	&	5.29	&	10.30 	\\
1705+7142	&	B	&	0.350	&	17	&	26	&	P19	&	24.75	&	24.93	&		&	4.01	&	3.04	&	6.37 	&	2.52	&	2.21	&	3.04 	\\
1706+36	&	B	&	0.918	&	15	&	12	&	P19	&	25.53	&	25.43	&		&	7.06	&	4.77	&	13.54 	&	3.68	&	3.05	&	4.77 	\\
1707+344	&	I	&	0.081	&	5	&	196	&	P19	&	22.95	&	24.54	&		&		&		&		&		&		&		\\
1713+504	&	B	&	1.090	&	21	&	39	&	P19	&	25.99	&	26.26	&		&	6.91	&	4.70	&	13.17 	&	3.63	&	3.02	&	4.70 	\\
1715+574A	&	B	&	0.599	&	14	&	21	&	P19	&	25.19	&	25.37	&		&	5.00	&	3.63	&	8.56 	&	2.93	&	2.51	&	3.63 	\\
1717+178	&	B	&	0.137	&	661	&	8.043	&	F11	&	25.52	&	23.61	&		&	23.69	&	12.58	&	68.05 	&	8.25	&	6.10	&	12.58 	\\
1731+325	&	B	&	0.375	&	37	&	187	&	P19	&	25.19	&	25.89	&		&	3.51	&	2.73	&	5.34 	&	2.31	&	2.05	&	2.73 	\\
1733+453	&	B	&	0.317	&	13	&	6	&	P19	&	24.60	&	24.27	&		&	5.30	&	3.79	&	9.23 	&	3.04	&	2.59	&	3.79 	\\
1738+1944	&	B	&	0.599	&	143	&	69	&	P19	&	26.24	&	25.92	&		&	11.54	&	7.08	&	26.09 	&	5.11	&	4.05	&	7.08 	\\
1738+476	&	B	&	0.950	&	848	&	1	&	F11	&	27.35	&	24.42	&		&	113.08	&	43.93	&	546.83 	&	23.38	&	14.91	&	43.93 	\\
1747+433	&	B	&	0.215	&	295	&	72	&	F11	&	25.55	&	24.94	&		&	10.07	&	6.35	&	21.75 	&	4.66	&	3.74	&	6.35 	\\
1749+701	&	B	&	0.770	&	1754	&	89.16	&	F11	&	27.59	&	26.30	&		&	42.60	&	20.12	&	148.79 	&	12.20	&	8.53	&	20.12 	\\
1750+374	&	B	&	0.599	&	28	&	5	&	P19	&	25.34	&	24.59	&		&	9.97	&	6.29	&	21.45 	&	4.63	&	3.72	&	6.29 	\\
1752+3212	&	B	&	0.599	&	35	&	7	&	P19	&	25.62	&	24.92	&		&	11.01	&	6.82	&	24.50 	&	4.95	&	3.94	&	6.82 	\\
1756+553	&	B	&	2.085	&	10	&	4	&	P19	&	26.12	&	25.72	&		&	11.51	&	7.06	&	25.98 	&	5.10	&	4.04	&	7.06 	\\
1757+703	&	B	&	0.407	&	11	&	1	&	P19	&	24.73	&	23.69	&		&	9.04	&	5.82	&	18.84 	&	4.34	&	3.52	&	5.82 	\\
1800+664	&	B	&	0.026	&	5	&	2	&	P19	&	21.96	&	21.56	&		&	1.53	&	1.41	&	1.77 	&	1.33	&	1.28	&	1.41 	\\
1807+698	&	B	&	0.051	&	1520	&	430	&	F11	&	24.99	&	24.44	&		&	7.36	&	4.94	&	14.32 	&	3.78	&	3.13	&	4.94 	\\
1811+442	&	B	&	0.350	&	6	&	22	&	P19	&	24.31	&	24.88	&		&	2.52	&	2.10	&	3.43 	&	1.85	&	1.70	&	2.10 	\\
1831+401	&	B	&	0.599	&	23	&	3	&	P19	&	25.62	&	24.73	&		&	12.47	&	7.53	&	28.93 	&	5.38	&	4.23	&	7.53 	\\
1831+559	&	B	&	0.599	&	13	&	5	&	P19	&	25.28	&	24.86	&		&	7.74	&	5.14	&	15.32 	&	3.91	&	3.22	&	5.14 	\\
1832+315	&	B	&	0.599	&	339	&	698	&	P19	&	26.65	&	26.96	&		&	9.24	&	5.92	&	19.39 	&	4.40	&	3.56	&	5.92 	\\
1832+474	&	II(G)	&	0.161	&	7.8	&	1445.28	&	F11	&	23.77	&	26.04	&		&		&		&		&		&		&		\\
1833+326	&	II(G)	&	0.058	&	160	&	1600	&	F11	&	24.14	&	25.14	&		&		&		&		&		&		&		\\
1836+171	&	I	&	0.017	&	17	&	2907.22	&	F11	&	22.06	&	24.29	&		&		&		&		&		&		&		\\
1839-48	&	I	&	0.112	&		&		&		&	24.98	&	25.87	&	Z95	&		&		&		&		&		&		\\
1841+317	&	B	&	0.448	&	42	&	100	&	P19	&	25.40	&	25.78	&		&	4.84	&	3.53	&	8.19 	&	2.86	&	2.46	&	3.53 	\\
1842+455	&	II(G)	&	0.092	&		&		&		&	23.76	&	25.73	&	F11	&		&		&		&		&		&		\\
1848+427	&	B	&	0.599	&	8	&	16	&	P19	&	24.89	&	25.19	&		&	3.98	&	3.02	&	6.31 	&	2.51	&	2.20	&	3.02 	\\
1853+671	&	B	&	0.212	&	12	&	1	&	P19	&	24.15	&	23.07	&		&	6.99	&	4.74	&	13.36 	&	3.65	&	3.04	&	4.74 	\\
1914-194	&	B	&	0.137	&	293	&	124	&	P19	&	25.16	&	24.78	&		&	7.10	&	4.80	&	13.64 	&	3.69	&	3.06	&	4.80 	\\
1915+3419	&	B	&	0.599	&	34	&	1	&	P19	&	25.59	&	24.06	&		&	18.97	&	10.53	&	50.60 	&	7.11	&	5.37	&	10.53 	\\
1926+611	&	B	&	0.369	&	825	&	1	&	F11	&	26.49	&	23.57	&		&	73.88	&	31.25	&	310.03 	&	17.61	&	11.69	&	31.25 	\\
1939+605	&	II(G)	&	0.201	&	28.54	&	1141.64	&	F11	&	24.56	&	26.16	&		&		&		&		&		&		&		\\
1954-55	&	I	&	0.060	&		&		&		&	23.91	&	25.60	&	Z95	&		&		&		&		&		&		\\
1957+4035	&	I	&	0.056	&	277	&	276	&	P19	&	24.34	&	24.34	&		&		&		&		&		&		&		\\
2003+454	&	B	&	0.599	&	557	&	110	&	P19	&	26.81	&	26.11	&		&	19.73	&	10.86	&	53.30 	&	7.30	&	5.50	&	10.86 	\\
2005-489	&	B	&	0.071	&	454	&	736	&	P19	&	24.75	&	24.96	&		&	3.95	&	3.00	&	6.25 	&	2.50	&	2.19	&	3.00 	\\
2007+777	&	B	&	0.342	&	823	&	50.28	&	PS93	&	26.36	&	25.15	&		&	22.30	&	11.98	&	62.76 	&	7.92	&	5.89	&	11.98 	\\
2010+723	&	B	&	0.369	&	1338	&	1	&	F11	&	26.72	&	23.59	&		&	94.84	&	38.16	&	432.51 	&	20.80	&	13.48	&	38.16 	\\
2013-308	&	I	&	0.089	&		&		&		&	23.03	&	24.51	&	Z95	&		&		&		&		&		&		\\
2020+6409	&	B	&	0.599	&	125	&	20	&	P19	&	26.10	&	25.30	&		&	14.86	&	8.66	&	36.53 	&	6.04	&	4.67	&	8.66 	\\
2021+317	&	B	&	0.599	&	3060	&	58	&	P19	&	27.50	&	25.78	&		&	54.31	&	24.43	&	205.66 	&	14.34	&	9.80	&	24.43 	\\
2022+542	&	B	&	0.599	&	991	&	100	&	P19	&	27.04	&	26.05	&		&	26.77	&	13.87	&	80.09 	&	8.95	&	6.54	&	13.87 	\\
2023+76	&	B	&	0.594	&	425	&	349	&	P19	&	26.46	&	26.38	&		&	11.02	&	6.82	&	24.51 	&	4.95	&	3.94	&	6.82 	\\
2028+1925	&	B	&	0.599	&	40	&	2	&	P19	&	25.65	&	24.35	&		&	16.78	&	9.55	&	42.96 	&	6.55	&	5.01	&	9.55 	\\
2029+121	&	B	&	1.215	&	1006	&	222	&	P19	&	27.51	&	26.85	&		&	26.71	&	13.85	&	79.85 	&	8.94	&	6.54	&	13.85 	\\
2030+547	&	B	&	1.262	&	943	&	54	&	P19	&	27.67	&	26.42	&		&	42.67	&	20.14	&	149.11 	&	12.21	&	8.54	&	20.14 	\\
2032+117	&	B	&	0.607	&	102	&	78	&	P19	&	26.13	&	26.02	&		&	9.58	&	6.10	&	20.36 	&	4.51	&	3.64	&	6.10 	\\
2048+0701	&	I	&	0.127	&		&		&		&	23.80	&	25.49	&	B11	&		&		&		&		&		&		\\
2052+0003	&	B	&	0.151	&	44	&	21	&	P19	&	24.40	&	24.08	&		&	4.76	&	3.49	&	8.02 	&	2.83	&	2.44	&	3.49 	\\
2053-201	&	I	&	0.156	&		&		&		&	23.79	&	25.47	&	Z95	&		&		&		&		&		&		\\
2058-282	&	I	&	0.040	&	63	&	1937	&	P19	&	23.37	&	24.85	&		&		&		&		&		&		&		\\
2104+763	&	II(G)	&	0.572	&	0.89	&	752.71	&	NED	&	24.10	&	27.02	&		&		&		&		&		&		&		\\
2104-256	&	I	&	0.039	&		&		&		&	22.94	&	24.48	&	Z95	&		&		&		&		&		&		\\
2116+203	&	B	&	1.680	&	110	&	101	&	P19	&	27.01	&	26.97	&		&	13.90	&	8.21	&	33.42 	&	5.78	&	4.50	&	8.21 	\\
2116+26	&	I	&	0.016	&		&		&		&	22.73	&	22.69	&	Z95	&		&		&		&		&		&		\\
2124+505	&	I	&	0.020	&	534	&	424	&	P19	&	23.90	&	23.80	&		&		&		&		&		&		&		\\
2131-021	&	B	&	1.285	&		&		&		&	27.60	&	25.60	&	F11	&	68.55	&	29.43	&	280.54 	&	16.75	&	11.20	&	29.43 	\\
2136-251	&	B	&	0.940	&	179	&	150	&	P19	&	26.94	&	26.87	&		&	13.82	&	8.17	&	33.16 	&	5.76	&	4.48	&	8.17 	\\
2141+279	&	II(G)	&	0.215	&	17.9	&	803.74	&	F11	&	24.41	&	26.06	&		&		&		&		&		&		&		\\
2143.4+0704	&	B	&	0.237	&	44.63	&	23.15	&	F11	&	24.85	&	24.56	&		&	5.76	&	4.06	&	10.32 	&	3.21	&	2.72	&	4.06 	\\
2144+147	&	B	&	0.599	&	23	&	1	&	P19	&	25.54	&	24.18	&		&	16.51	&	9.42	&	42.03 	&	6.48	&	4.96	&	9.42 	\\
2149+17	&	B	&	0.871	&	648	&	372	&	P19	&	27.14	&	26.90	&		&	16.96	&	9.63	&	43.59 	&	6.60	&	5.04	&	9.63 	\\
2152-69	&	I	&	0.027	&		&		&		&	24.11	&	25.56	&	Z95	&		&		&		&		&		&		\\
2153+377	&	II(G)	&	0.290	&	9	&	2565.2	&	F11	&	24.44	&	26.89	&		&		&		&		&		&		&		\\
2155-304	&	B	&	0.117	&	350	&	218.23	&	F11	&	25.08	&	24.87	&		&	6.10	&	4.25	&	11.15 	&	3.34	&	2.81	&	4.25 	\\
2158-206	&	B	&	0.370	&	87	&	23	&	P19	&	25.69	&	25.11	&		&	10.51	&	6.56	&	23.01 	&	4.80	&	3.83	&	6.56 	\\
2200+420	&	B	&	0.069	&	1990	&	11.03	&	K10	&	25.36	&	23.11	&		&	27.60	&	14.22	&	83.42 	&	9.13	&	6.66	&	14.22 	\\
2201+044	&	B	&	0.028	&		&		&		&	23.41	&	24.00	&	F11	&	1.60	&	1.46	&	1.88 	&	1.37	&	1.31	&	1.46 	\\
2208+457	&	B	&	0.599	&	39	&	1	&	P19	&	25.64	&	24.05	&		&	20.19	&	11.07	&	54.98 	&	7.42	&	5.57	&	11.07 	\\
2209+236	&	B	&	1.125	&	430	&	1	&	P19	&	27.45	&	24.82	&		&	97.55	&	39.03	&	449.04 	&	21.19	&	13.70	&	39.03 	\\
2213+287	&	B	&	0.229	&	103	&	74	&	P19	&	25.25	&	25.11	&		&	6.38	&	4.40	&	11.83 	&	3.44	&	2.88	&	4.40 	\\
2214+1350	&	I	&	0.026	&		&		&		&	21.50	&	24.10	&	B11	&		&		&		&		&		&		\\
2214+241	&	B	&	0.505	&	420	&	265	&	P19	&	26.44	&	26.24	&		&	11.77	&	7.19	&	26.77 	&	5.17	&	4.09	&	7.19 	\\
2223-052	&	B	&	1.404	&		&		&		&	28.70	&	27.20	&	F11	&	83.56	&	34.48	&	365.31 	&	19.11	&	12.54	&	34.48 	\\
2225+692	&	B	&	0.599	&	27	&	67	&	P19	&	25.60	&	26.00	&		&	5.27	&	3.78	&	9.17 	&	3.03	&	2.58	&	3.78 	\\
2231+3921	&	I	&	0.018	&		&		&		&	22.40	&	23.57	&	B11	&		&		&		&		&		&		\\
2243+394	&	II(G)	&	0.081	&		&		&		&	24.23	&	25.47	&	F11	&		&		&		&		&		&		\\
2251+006	&	B	&	0.939	&	244	&	175	&	P19	&	26.88	&	26.74	&		&	14.03	&	8.27	&	33.82 	&	5.82	&	4.52	&	8.27 	\\
2251+244	&	B	&	2.328	&	134	&	666	&	P19	&	27.50	&	28.19	&		&	10.77	&	6.70	&	23.79 	&	4.88	&	3.89	&	6.70 	\\
2254+074	&	B	&	0.190	&	454	&	14.54	&	F11	&	25.64	&	24.15	&		&	18.96	&	10.53	&	50.55 	&	7.11	&	5.37	&	10.53 	\\
2309+184	&	II(G)	&	0.428	&	7	&	744.4	&	F11	&	24.72	&	26.75	&		&		&		&		&		&		&		\\
2313+147	&	B	&	0.163	&	26	&	1	&	P19	&	24.27	&	22.86	&		&	9.29	&	5.95	&	19.53 	&	4.42	&	3.57	&	5.95 	\\
2315+115	&	B	&	0.567	&	13	&	16	&	P19	&	25.14	&	25.23	&		&	5.16	&	3.71	&	8.91 	&	2.98	&	2.55	&	3.71 	\\
2318+235	&	II(G)	&	0.268	&	21	&	369	&	P19	&	24.72	&	25.96	&		&		&		&		&		&		&		\\
2320+0813	&	I	&	0.011	&		&		&		&	21.60	&	22.77	&	B11	&		&		&		&		&		&		\\
2320+343	&	B	&	0.098	&	30	&	48	&	P19	&	23.86	&	24.07	&		&	2.58	&	2.13	&	3.54 	&	1.88	&	1.72	&	2.13 	\\
2320+417	&	B	&	0.152	&	7	&	39	&	P19	&	23.64	&	24.39	&		&	1.62	&	1.47	&	1.90 	&	1.38	&	1.32	&	1.47 	\\
2322-123	&	I	&	0.082	&	12	&	380	&	P19	&	23.33	&	24.83	&		&		&		&		&		&		&		\\
2326+174	&	B	&	0.213	&	18	&	8	&	P19	&	24.33	&	23.98	&		&	4.70	&	3.45	&	7.87 	&	2.81	&	2.42	&	3.45 	\\
2329+3433	&	B	&	0.599	&	25	&	3	&	P19	&	25.56	&	24.64	&		&	12.42	&	7.50	&	28.75 	&	5.36	&	4.22	&	7.50 	\\
2335+267	&	I	&	0.030	&		&		&		&	23.37	&	24.84	&	F11	&		&		&		&		&		&		\\
2335+358	&	B	&	2.280	&	29	&	96	&	P19	&	26.69	&	27.21	&		&	8.21	&	5.39	&	16.57 	&	4.07	&	3.33	&	5.39 	\\
2338+2701	&	I	&	0.030	&		&		&		&	23.70	&	24.54	&	B11	&		&		&		&		&		&		\\
2343-151	&	B	&	0.224	&	8	&	1	&	P19	&	24.04	&	23.14	&		&	5.89	&	4.13	&	10.64 	&	3.26	&	2.75	&	4.13 	\\
2347+1924	&	B	&	0.515	&	3	&	1	&	P19	&	24.38	&	23.91	&		&	5.24	&	3.76	&	9.10 	&	3.02	&	2.58	&	3.76 	\\
2354-113	&	B	&	0.949	&	12	&	276	&	P19	&	25.52	&	26.88	&		&	2.66	&	2.19	&	3.68 	&	1.92	&	1.75	&	2.19 	\\

\hline

\end{longtable}


\end{document}